%% file: revised_new_manuscript.tex
\documentclass{arxiv}
\usepackage{graphicx}
\usepackage{xcolor}
\usepackage[normalem]{ulem}
\usepackage{newtxtext}
\usepackage{subcaption}
\usepackage{newtxmath}
\usepackage{soul}
\usepackage{natbib}
\usepackage{hyperref}
\hypersetup{
    colorlinks = true,
    urlcolor   = blue,
    citecolor  = black,
}
\input{definitions.tex}

\newcommand{\RomanNumeralCaps}[1]

\title{Scale-by-scale energy transfers in bubbly flows}
\author{Hridey Narula\aff{1}\corresp{\email{hrideynarula@tifrh.res.in}},
  Vikash Pandey \aff{2},
  Dhrubaditya Mitra \aff{2,3},
  Prasad Perlekar\aff{1}\corresp{\email{perlekar@tifrh.res.in}}}
\affiliation{
\aff{1}Tata Institute of Fundamental Research,
Gopanpally, Hyderabad 500046, India
\aff{2} Nordita, KTH Royal Institute of Technology 
and Stockholm University, 
 Hannes Alfvéns väg 12, 11419 Stockholm, Sweden
\aff{3} Digital Futures, Stockholm, Sweden.  
}

\begin{document}
\maketitle

\begin{abstract}
Buoyancy-driven bubbly flows naturally have spatially-dependent density fields, which allow for multiple definitions of the scale-dependent (or filtered) energy.
A priori, it is not obvious which of these provide the most physically apt scale-by-scale budget. In the present study, we compare two such definitions, based on (a) filtered momentum and filtered velocity \citep{Pandey_Ramadugu_Perlekar_2020}, and (b) Favre filtered energy \citep{Aluie2013,PandeyMitraPerlekar2023}. We also derive a K\'arm\'an-Howarth-Monin (KHM) relation using the momentum-velocity correlation function and contrast it with the scale-by-scale energy budget obtained in (a). 
We find that for the volume fraction and Atwood number explored, irrespective of the definition, energy transfers due to the advective nonlinearity and surface tension are identical.
However, discrepancies arise for the buoyancy and pressure contributions.
We show that the Favre filtered definition is the more appropriate choice, within which buoyancy injects energy, pressure transfers energy to large scales, and both advective nonlinearity and surface tension transfer energy downscales where it is dissipated by viscosity.
\end{abstract}

\begin{keywords}
bubble dynamics, multiphase flow, turbulence simulation
\end{keywords}

\section{Introduction}
The presence of bubbles is known to dramatically alter transport properties in a variety of natural and industrial flows \citep{clift2005bubbles,Balchandar_2010}. Perhaps the simplest, yet not very well understood, realization of such a phenomenon are complex spatio-temporal flows (often referred to as pseudo-turbulence or bubble-induced agitation) generated by a dilute swarm of bubbles rising through an otherwise quiescent fluid \citep{LanceBataille1991,Prakash_2016,Risso_2018,ChaoSun_2020,RuiNi2024}. The wake flows, generated by individual bubbles in the swarm, agitate the ambient fluid and interact with each other to generate pseudo-turbulence.

The key dimensionless numbers that characterize pseudo-turbulence are the Galilei number
$\Ga=\sqrt{\rho_L (\rho_L-\rho_B)gD^3/\mu^2}$ (the ratio of buoyancy to viscous forces), the Bond number $\Bo=(\rho_L-\rho_B)g D^2/\sigma$ (the ratio of buoyancy to surface tension forces), and the Atwood number $\At=(\rho_L-\rho_B)/(\rho_L+\rho_B)$, where $\rho_L$ is the density of the liquid phase, $\rho_B$ is the density of bubble, $\mu$ is the viscosity of the liquid phase, $g$ is the gravitational acceleration, and $D$ is the bubble diameter. Experiments on bubbly flows are typically conducted in the dilute regime, characterized by a bubble volume fraction $\phi<2\%$, where interactions between bubbles can be ignored. 
Furthermore, for experimentally relevant $\Ga<1000$ and $\Bo<3$, the bubbles are roughly  ellipsoidal and they do not undergo breakup.

Similar to classical turbulence, it is usual to characterize bubbly flows using the energy spectrum. Experimental and numerical studies at moderate $\Ga\,(<1000)$ and varying
$\At$ characterize pseudo-turbulence by a power-law scaling of the energy
spectrum $E(k)\sim k^{-\beta}$, where the exponent $\beta$ depends on
$\Ga$ and $\At$ \citep{LanceBataille1991,ESMAEELI_TRYGGVASON_1998,
  ESMAEELI_TRYGGVASON_1999,RIBOUX_2010,MARTINEZ_2010,Prakash_2016,
  Pandey_Ramadugu_Perlekar_2020, Innocenti_Jaccod_Popinet_Chibbaro_2021,
  Ma_Hessenkemper_Lucas_Bragg_2022, Pandey_Mitra_Perlekar_2022,
  Ravisankar_Zenit_2024}.  
In a recent study, using high-resolution numerical simulations at large
$\Ga\,(>1000)$, \citet{PandeyMitraPerlekar2023} show that Kolmogorov
turbulence coexists with pseudo-turbulence.  This has been experimentally corroborated by \citet{ma2025}.\newline 
There are two salient features of incompressible, single-phase homogeneous,
and isotropic turbulence \citep{Frisch_1995}: power-law scaling of the energy
spectrum $E(k)\sim k^{-5/3}$, and the cascade of energy from large to small scales
in the inertial range while maintaining constant energy flux. These are now well-established by state-of-the-art numerical simulations and experiments \citep{Kaneda_2016,Iyer2020,Bodenschatz2023}.\newline 
What are the mechanisms of inter-scale energy transfers in bubbly flows? 
Different studies have attempted to address this question using a scale-by-scale budget equation \citep{Pandey_Ramadugu_Perlekar_2020,
  PandeyMitraPerlekar2023, RAMIREZ2024104860}. However, as is the case for variable density flows, there is no unique way to define scale-dependent kinetic energy, making the analysis
ambiguous \citep{Aluie2013,Eyink_2018,ZhaoAluie_2018}. Some studies \citep{Prakash_2016, Innocenti_Jaccod_Popinet_Chibbaro_2021,
  Ma_Hessenkemper_Lucas_Bragg_2022} avoid this ambiguity by focusing only on
the liquid phase, where the density is constant.
This, of course, misses the correlations between points that are situated either within the bubble phase or across it, both of which are crucial to the energy budget.\newline 
The ambiguity in the definition of scale-dependent energy has been addressed mostly in the context of compressible turbulence \citep{Aluie2013,Wang_2013, Eyink_2018,Hellinger_2020,Hellinger_2021b,Hellinger_2021a}. Notably \citet{Aluie2013, Eyink_2018} argue that for compressible turbulent flows, the Favre definition (see section \ref{subsec:favre}) of the scale-dependent energy is the most appropriate.\newline 
In contrast to compressible turbulence, bubbly flows are, to a good approximation, incompressible and the experimentally accessible Reynolds numbers are not too large.
Thus, it behooves us to understand and investigate the scale-by-scale energy transfer using different definitions of scale-dependent energy that have been used in the literature.
In particular, we follow the procedure outlined in \citet{Frisch_1995,Pope_2000,Aluie2013} and
derive the scale-by-scale budget equation for two different definitions of scale-dependent (or filtered) energy.
For one of these definitions, we also derive an equivalent
K\'arm\'an-Howarth-Monin (KHM) relation to study kinetic energy transfers in terms of spatial correlations  \citep{GaltierBanerjee2011, Fabien_2025}. 
We show that although different definitions show a net transfer of energy from large to small scales, the individual contributions to the inter-scale energy transfer do not have identical physical interpretation.
However, if we impose the physically intuitive constraints that the buoyancy injection should happen primarily inside the bubble and it should be bounded above by the unfiltered energy injection, then the Favre definition is found to be more appropriate.\newline 
The rest of the paper is organized as follows: in Section \ref{sec:model} we discuss the equations of motion and numerical details, next we provide an overview of different methodologies and definitions that can be used to study these flows in Section \ref{sec:ScaleByScale}.
The results are presented in Section \ref{sec:Results}.
\section{Model}\label{sec:model}
We study bubbly flows by using the Navier-Stokes equations, combined with surface tension and buoyancy forces \citep{BUNNER_TRYGGVASON_2002, Pandey_Ramadugu_Perlekar_2020},
\begin{subequations}
\begin{align}
 \text{D}_t \rho \bu &= -\bnabla p + \mu \nabla^{2} \bu + \boldsymbol{F}^g + \boldsymbol{F}^\sigma, \label{eq:ns1} \\
 \text{D}_t \rho &= 0, \label{eq:ns2}\\
 \bnabla   \cdot \bu & = 0.
 \label{eq:ns3}
\end{align}
\label{eq:ns}
\end{subequations}
Here $\bu$, $\rho$ and $p$ are, respectively, the hydrodynamic velocity field, the density field, and the pressure field at position $\bx = (x, y, z)$ and time $t$.  $\text{D}_t=\partial_t + \bu\cdot\nabla$ is the material derivative. The dynamical viscosity $\mu$ is assumed to be identical for both the phases.
The surface tension force  $\boldsymbol{F}^\sigma = \sigma \kappa \boldsymbol{n}$, where $\sigma$ is the surface tension coefficient of the bubble-fluid interface, $\kappa$ is the curvature (field) and $\boldsymbol{n}$ is the normal to the bubble interface.
The buoyancy force  $\boldsymbol{F}^g = (\rho-\rho_a)\boldsymbol{g}$, where $\boldsymbol{g}$ is the gravitational acceleration (pointing downward), $\rho_a=\rho_L+\alpha (\rho_B-\rho_L)$ is the average density and $\alpha$ is the volume fraction of the bubble phase. The second term in $\boldsymbol{F}^g$ prevents uniform downward acceleration of the velocity field and ensures zero average momentum at all times.

\subsection{Direct Numerical Simulations \label{sec:DNS}}
We perform a direct numerical simulation of \eqref{eq:ns} in a periodic cubic domain of volume $V_{\rm Box}$ discretized with $N^3$ equi-spaced grid points.
A summary of the parameters used in our study is provided in table~\ref{tab:dataset}. 
In the small $\At$ number regime ({\tt run R1}), we use the Boussinesq approximation wherein the density variation is only retained in the buoyancy force term and solve \eqref{eq:ns1} using  a pseudo-spectral  method. 
In the high $\At$ number regime ({\tt run R2}) we numerically integrate \eqref{eq:ns1} using the finite-volume solver PARIS \citep{Aniszewski2021}.Using \eqref{eq:ns1} and \eqref{eq:ns3}, we get a non-separable elliptic equation $\nabla \cdot (\rho^{-1} \nabla p)=S$ for the pressure field, where the source $S$ is obtained from \eqref{eq:ns1}.
This equation is numerically solved using a multigrid scheme to obtain the  pressure field \citep{Aniszewski2021}.

Bubble fronts are evolved using a front-tracking algorithm \citep{Aniszewski2021}. Initially, the $N_B$ spherical bubbles of diameter $D$ are randomly distributed to achieve the volume fraction $\alpha\equiv N_BV_B/V_{\rm{Box}}$, where $V_B=\pi D^3/6$ is the volume of a single bubble.
From the bubble front, an indicator field $\phi$ is obtained on the grid by numerically solving $\nabla^2 \phi=\nabla \cdot \boldsymbol{n}$. 
Depending on the numerical approximation used to extrapolate the normal vector on to the grid, we obtain a $\phi$ that smoothly varies from $1$ (inside the bubble) to $0$ (in the liquid) over two to three points on the grid.
The density field expressed as $\rho=\phi (\rho_B-\rho_L)+\rho_L$ satisfies \eqref{eq:ns2}.
We refer the reader to \citet{Pandey_Ramadugu_Perlekar_2020,Aniszewski2021} for a more  detailed discussion of the numerical implementation.

\begin{table}
  \begin{center}
\def~{\hphantom{0}}
  \begin{tabular}{lcccccccc}
Runs & $N$ & $V_{{\rm box}}$  & $N_B$ & $\alpha$ & Bo & At & Ga & Source \\[3pt]
\tt R1 & 512  & $(2\upi)^3$& 12&3.2\% & 1.75 & 0.04 & 605 & {\tt R4} of \citep{PandeyMitraPerlekar2023} \\
\tt R2 & 504 & $(2\upi)^3$ & 12 & 3.2\% & 1.75 & 0.8 & 1059 & {\tt R8} of \citep{PandeyMitraPerlekar2023} \\
  \end{tabular}
  \caption{Table of parameters used in our DNS. We use a pseudo-spectral flow solver for {\tt R1} and PARIS for {\tt R2}. The bubble evolution is done using front-tracking in both the cases.
  For run {\tt R1} we use a pseudo-spectral flow solver, and for run {\tt R2} we use the finite-difference based PARIS flow solver \citep{Aniszewski2021}. The bubble evolution is done using front-tracking for both the runs. The bubble diameter $D=1.08$ for both the runs.}
  \label{tab:dataset}
  \end{center}
\end{table}

\subsection{Characterization of the Steady-State}
\begin{figure}
\centering
  
  \includegraphics[width=0.322\linewidth]{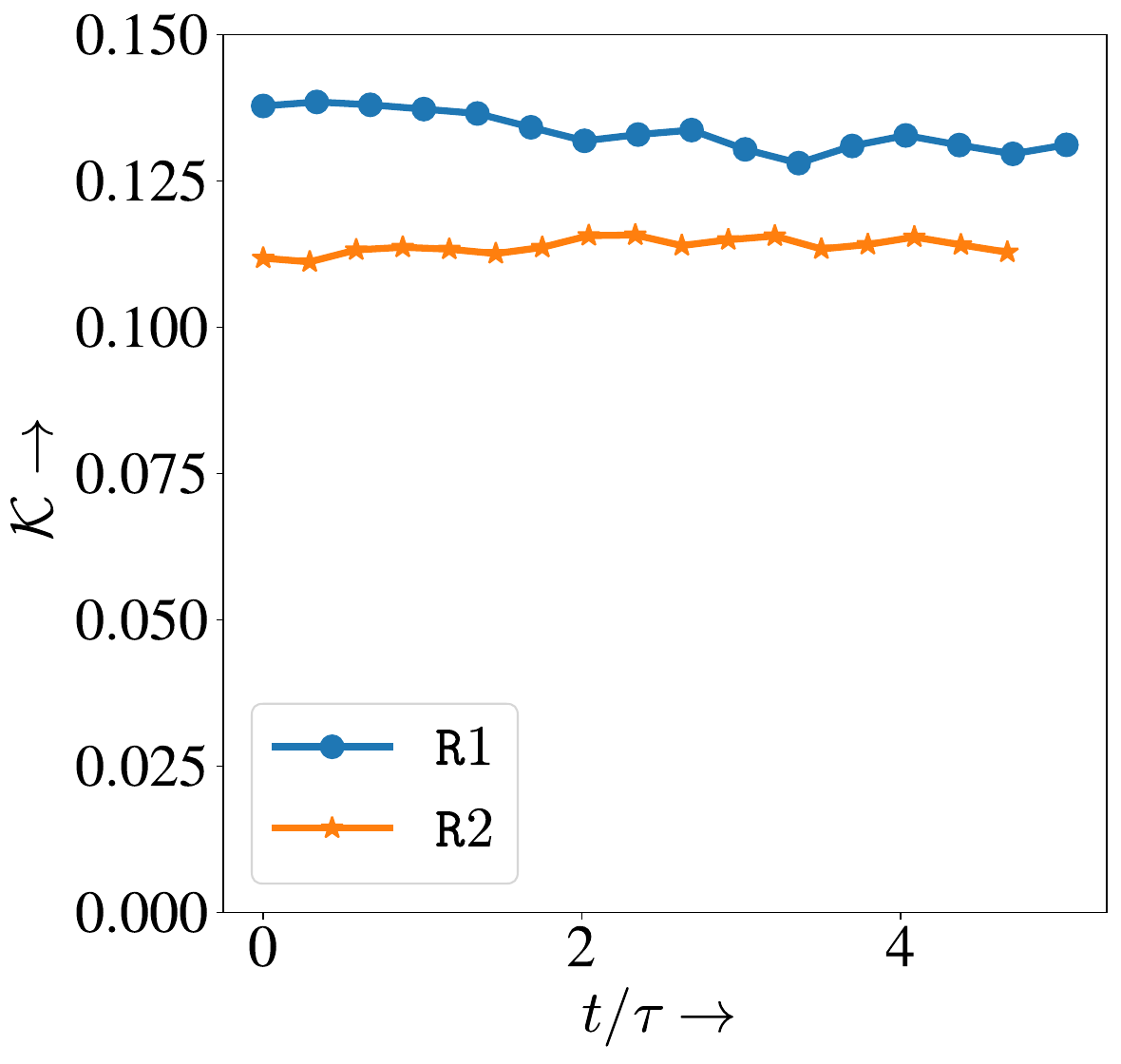}
  \includegraphics[width=0.315\linewidth]{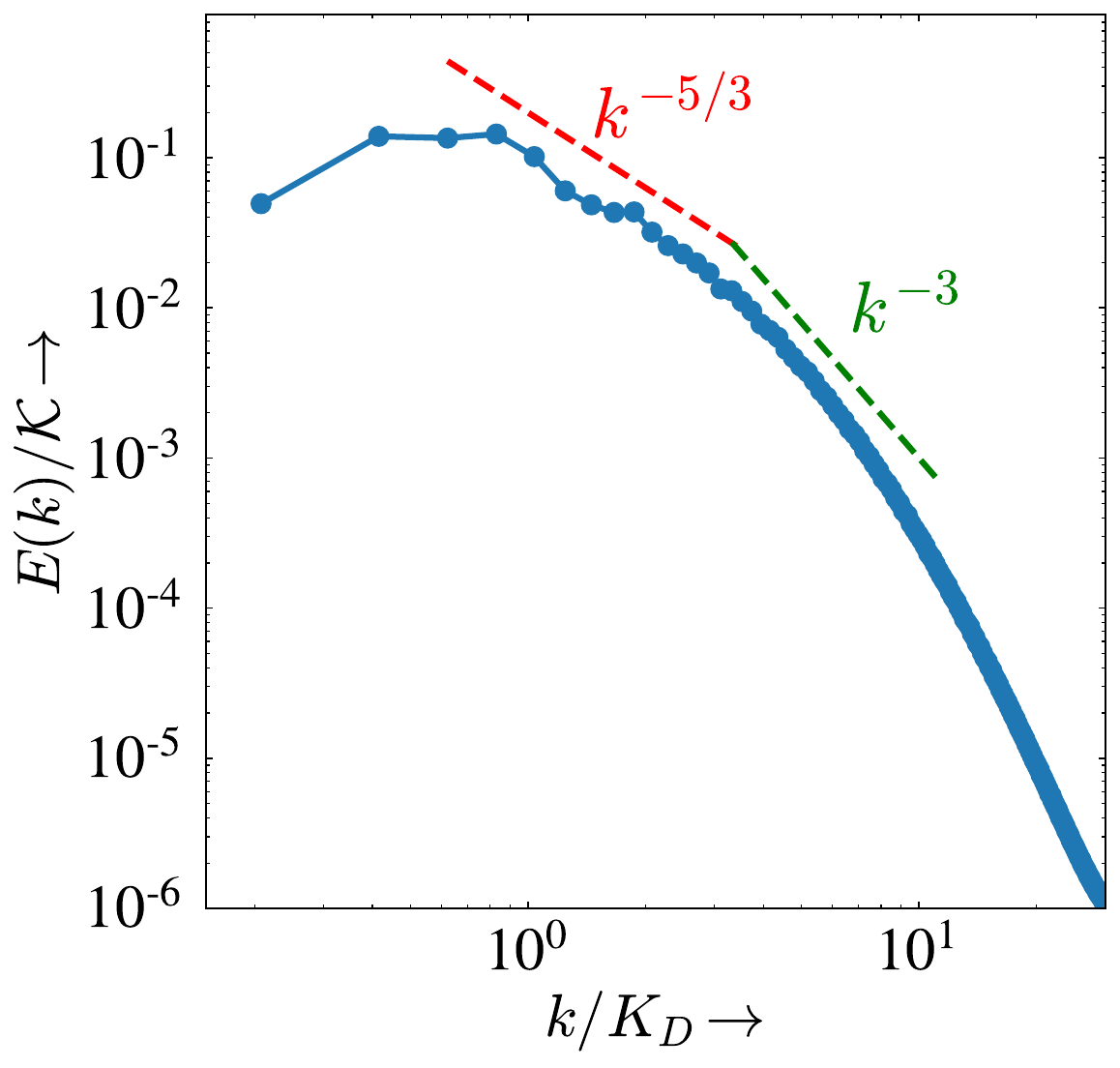}
  \includegraphics[width=0.34\linewidth]{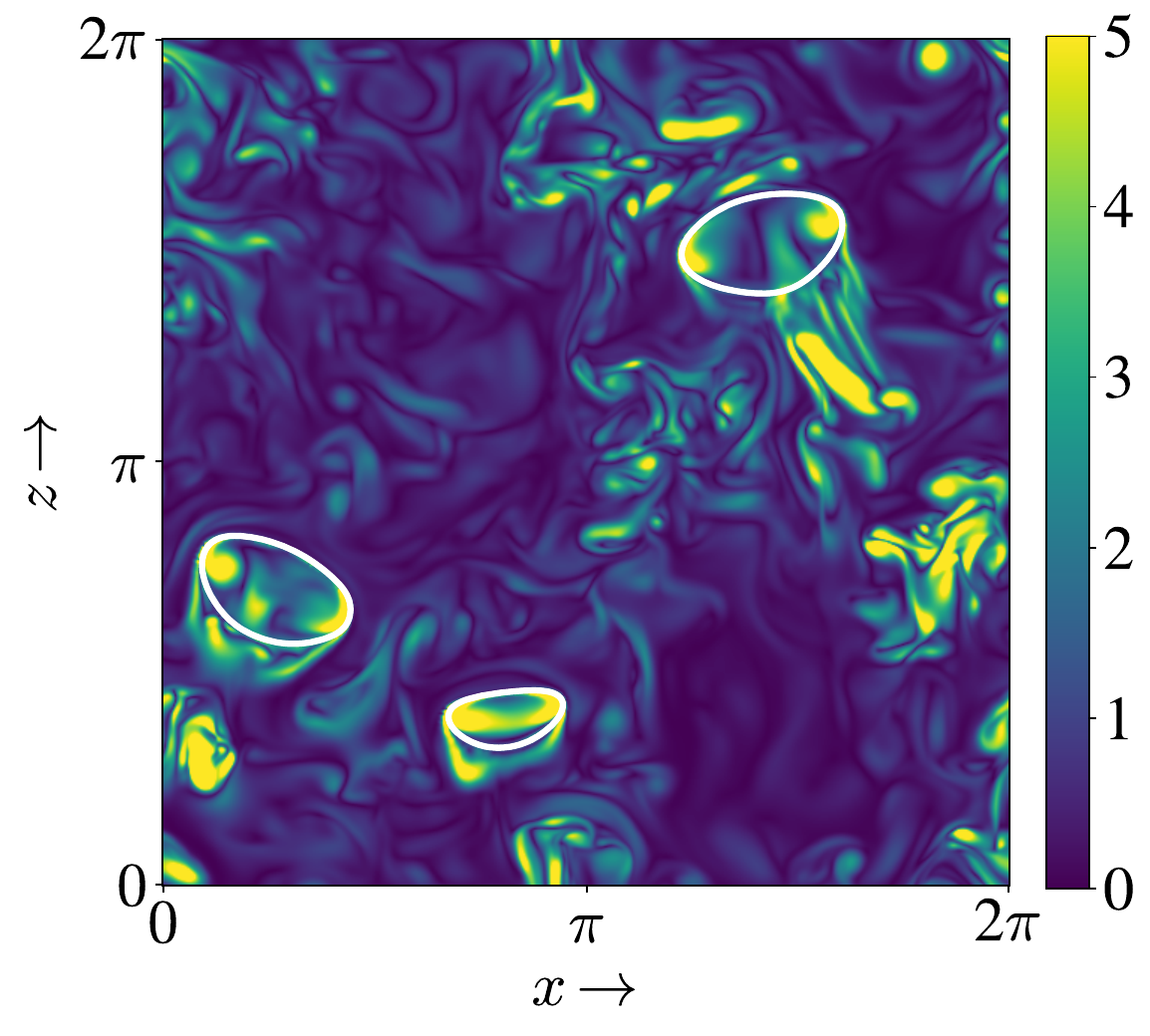}

\caption{(a) 
 Kinetic energy time series in the statistically steady state. (b) (time-averaged) energy spectrum $E(k)$ (normalized by the unfiltered kinetic energy) in the statistically steady state for run {\tt R2}.
 We indicate the Kolmogorov ($k^{-5/3}$) and the pseudo-turbulence ($k^{-3}$) scaling range.
 The $k^{-5/3}$ scaling becomes more prominent for large Galilei number ($\Ga\ge1000)$ \citep{PandeyMitraPerlekar2023,ma2025}.
 $K_D=2\upi/D$ is the wavenumber corresponding to the bubble diameter $D$ and $\tau$ is the large eddy turnover time. (c) pseudo-color plot of the magnitude of the vorticity field $|\boldsymbol{\omega}|$ in a representative 2$d$ slice containing bubbles for run {\tt R2}  (the bubble-liquid interface is marked in white).}\label{fig:hga_r8_representative}
\end{figure}

We evolve \eqref{eq:ns} for about $5\tau$, where $\tau\equiv {\mathcal L}/u_{z,{\rm rms}}$ is the integral time scale, 
$\mathcal{L}=3\upi/4 (\sum_k E(k)/k)/\sum_kE(k)$ is the integral length scale, $u_{z,{\rm rms}}$ is the rms value of the vertical velocity, and $E(k)$ is the energy spectrum, $E(k)\equiv \sum_{k-1/2<|\boldsymbol{q}|<k+1/2} |\widehat{\bu}(\boldsymbol{q})|^2$  with $\widehat{\boldsymbol{u}}(\boldsymbol{q})$ being the Fourier transform of the velocity field.
In figure \ref{fig:hga_r8_representative} (a, b), we show the time evolution of the kinetic energy ${\mathcal K}=1/V_{\rm Box} \int  1/2\,\rho \bu^2 {\rm d}V_{\rm Box}$ in the statistically steady state, and the steady state energy spectrum $E(k)$ showing the Kolmogorov ($-5/3$) and pseudo-turbulence ($-3$) scaling regimes \citep{PandeyMitraPerlekar2023}.
Figure~\ref{fig:hga_r8_representative} (c) shows a representative pseudo-color visualization of a 2D slice of the vorticity magnitude field, superimposed with the bubble interface.
Observe the pronounced wake flows produced by bubbles and their interactions with trailing bubbles and the wakes of neighboring bubbles.
In the subsequent sections, for the runs ${\tt R1}$ and ${\tt R2}$, statistical analysis is performed using fifteen independent realizations of the flow separated by approximately $0.3 \tau$.

\section{Scale-by-Scale Energy Transfers\label{sec:ScaleByScale}}
For any field $\psi({\bf x})$, we define the corresponding filtered field as
\begin{equation}\label{eq:filter_defn}
    \overline{\psi}_K (\boldsymbol{x}) = G_r * \psi = \int {\rm d}^3 \boldsymbol{k}\, G_K(\boldsymbol{k})\, \hat{\psi}(\boldsymbol{k})\, e^{i\boldsymbol{k}\cdot\boldsymbol{x}},
\end{equation}
where $G_r$ is the real space filtering kernel, $*$ denotes convolution, $G_K$ is a smooth low-pass filter in Fourier space that suppresses fluctuations at scales $<1/K$, $\widehat{\psi}$ is the Fourier transform of $\psi$, and  $r \sim 1/K$ is the filtering length scale corresponding to the filtering wavenumber $K$ \citep{Pope_2000,DomaradzkiCarati_2007,Sadek2018}.
In what follows, we will always use an isotropic Gaussian filter, $G_K(\boldsymbol{k}) = \exp(-\upi^2k^2/24K^2)$, which is smooth and positive-definite in both real space and Fourier space.

Some commonly used definitions of the scale-dependent (filtered) kinetic energy used for studying variable density or compressible flows are ($\langle\rangle$ denote spatial averaging):

\begin{itemize}
\item [{\tt F1}] : $\mathcal{E}_K \equiv \frac{1}{2}\langle\overline{\rho  \bu }_K \cdot \overline{\bu}_K\rangle$ \citep{Graham_2010,Salvese_2013,Pandey_Ramadugu_Perlekar_2020},
\item [{\tt F2}]: $\mathcal{E}_K \equiv \frac{1}{2} \langle \overline{\rho\bu}_K\cdot\overline{\rho\bu}_K/\overline{\rho}_K\rangle$ \citep{Favre_1965,Aluie_2011,PandeyMitraPerlekar2023},
\item [{\tt F3}]: $\mathcal{E}_K \equiv \frac{1}{2}\langle\overline{\rho}_K\,\overline{\bu}_K \cdot \overline{\bu}_K\rangle$ \citep{CHassaing1985}, and, 
\item [{\tt F4}]: $\mathcal{E}_K \equiv \frac{1}{2} \langle (\overline{\sqrt{\rho}\bu}_K)^2\rangle$ \citep{Kida_1990,Miura_Kida_1995,Girimaji_2019,SchmidtGrete_2019}.
\end{itemize}

Note that with the Gaussian filter, except for {\tt F1}, the scale-dependent kinetic energy is point-wise positive. All of the above definitions are identical for flows with uniform density $\rho=\rho_0$.
Besides filtering, inter-scale energy transfers can also be investigated from the time evolution of appropriate two-point correlators, that is, the K\'arm\'an-Howarth-Monin relation \citep{Frisch_1995}.
Two commonly used correlators for variable density flows are:
\begin{itemize}
\item [{\tt C1}]: $\mathcal{R}(\boldsymbol{r}) = \frac{1}{4}\langle \rho(\boldsymbol{x}) \bu(\boldsymbol{x})\cdot \bu(\boldsymbol{x+r})+\rho(\boldsymbol{x}) \bu(\boldsymbol{x})\cdot \bu(\boldsymbol{x-r})\rangle$ \citep{GaltierBanerjee2011,BanerjeeKritsuk_2017,Fabien_2025} and,
\item [{\tt C2}]: $\mathcal{R}(\boldsymbol{r}) = \frac{1}{2}\langle \sqrt{\rho\bu}(\boldsymbol{x}) \cdot \sqrt{\rho\bu}(\boldsymbol{x+r})\rangle$ \citep{Wang_2013,Hellinger_2021b},
\end{itemize}
where $\boldsymbol{r}$ is the increment vector.
Some studies have previously looked at the correlation tensor, $R_{ij}(\boldsymbol{r}) = 1/2\,\langle \rho(\boldsymbol{x}) u_i(\boldsymbol{x})u_j(\boldsymbol{x+r}) \rangle$, \citep{HILL_2001,Arun_Sameen_Srinivasan_Girimaji_2021,Hamba_2022}, and identified the scale-dependent energy as the trace of the shell-averaged correlation tensor.

The filtering and correlation function approaches are connected to each other. Using the Wiener-Khinchin theorem \citep{kubo1991statistical}, we obtain the following relation between the filtered energy ${\rm d}\mathcal{E}_K/{\rm d}K$ (with a sharp spectral filter as in \citet{Frisch_1995}) and the shell-averaged correlator $R(r)$\footnote{We sketch out a proof of this relation in Appendix \ref{sec:mapping_khm}}:

\begin{equation}\label{eq:wienerkhinchin_L2}
    R(r) = \int {\rm d}K \frac{{\rm{d}\mathcal{E}_K}}{{\rm d}K} \frac{\sin(Kr)}{Kr}.
\end{equation}
Equation \eqref{eq:wienerkhinchin_L2} connects the {\tt C1} and {\tt C2} schemes to their filtering analogs {\tt F1} and {\tt F4}, respectively.
For {\tt F2} and {\tt F3} there is no such mapping because the scale-dependent energy is not quadratic in the filtered fields.

We now summarize some choices that have been used in the past to study energy transfers in bubbly flows.
Motivated by \citet{GaltierBanerjee2011}, \citet{Pandey_Ramadugu_Perlekar_2020,Ramadugu_Pandey_Perlekar_2020} defined the filtered energy as per definition {\tt F1}, corresponding to the velocity-momentum correlator ({\tt C1}).
For bubbly flows, they show that the buoyancy contribution varies non-monotonically with respect to the filter cutoff $K$, and the total energy transfer (via advective nonlinearity, pressure and surface tension) is from large scales to small scales.
In a subsequent work, \citet{PandeyMitraPerlekar2023} used density-weighted Favre velocity to study the energy transfers, i.e., definition {\tt F2}.
This approach is extensively used to study scale-by-scale budget in compressible turbulence \citep{Aluie2013}, and leads to a (point-wise) positive definite filtered energy.
\citet{PandeyMitraPerlekar2023} found that the buoyancy contribution thus obtained is a monotonically increasing function of the filtering wavenumber $K$. 
While the pressure contribution here transfers energy from small to large scales, the net transfer is still downscales.
Alternatively, \citet{RAMIREZ2024104860} circumvented the need to define filtered energy by writing a ``\textit{spectral balance}" equation instead. 
However, this comes at the cost of containing an additional inertial time-derivative term in the budget, which has a non-zero contribution in the statistically steady state, and thus has to be interpreted carefully.

In the following, we review the scale-by-scale energy budget equations obtained using definitions {\tt F1} and {\tt F2}. For {\tt F1}, we derive an equivalent KHM relation for the velocity-momentum correlation using the definition {\tt C1}.

\subsection{Scale-by-scale budget for $\mathcal{E}_K=(1/2)\langle  \overline{\rho \bu}_K \cdot \overline{\bu}_K\rangle$ ({\tt F1}) \label{sec:f1}}
We first derive the scale-by-scale budget using the filtering approach presented in \citet{Pandey_Ramadugu_Perlekar_2020}. Starting from the Navier-Stokes equations, we obtain the following equations for the filtered momentum,
\begin{align}
 \partial_t \overline{\rho \bu}_K +\bnabla \cdot \overline{\rho\bu\bu}_K = -\bnabla\, \overline{p}_K + \mu \nabla^{2} \overline{\bu}_K + \overline{\boldsymbol{F}^g}_K+ \overline{\boldsymbol{F}^\sigma}_K,
 \label{eq:film}
 \end{align}
and the filtered velocity,
\begin{align}
 \partial_t \overline{\bu}_K +\bnabla \cdot \overline{\bu \bu}_K = -\overline{\left(\frac{\bnabla p}{\rho}\right)}_K
  + \mu \overline{\left(\frac{\nabla^{2} \bu}{\rho}\right)}_K + \overline{\left(\frac{\boldsymbol{F}^g}{\rho}\right)}_K+ \overline{\left(\frac{\boldsymbol{F}^\sigma}{\rho}\right)}_K.
 \label{eq:filu}
 \end{align}
Taking the dot product of \eqref{eq:film} with  $\overline{\bu}_K$, \eqref{eq:filu} with $\overline{\rho \bu}_K$, and averaging the two we obtain the following equation for the filtered energy $\mathcal{E}_K$:
\begin{equation}\label{eq:sbs_generic}
    \partial_t \mathcal{E}_K = \mathcal{N}_K + \mathcal{D}_K + \mathcal{P}_K + \mathcal{F}^g_K + \mathcal{F}^\sigma_K,
\end{equation}
where $\mathcal{N}_K$ is the advective nonlinear contribution, $\mathcal{D}_K$ is the viscous dissipation, $\mathcal{P}_K$ is the pressure contribution, $\mathcal{F}^g_K$ is the buoyancy injection and $\mathcal{F}^\sigma_K$ is the surface tension contribution.
The explicit expressions for each of these contributions are given below,
\begin{equation}\label{eq:f1_contri}
\begin{aligned}
    \mathcal{N}_K &= -\frac{1}{2} \left\langle  
        \overline{\boldsymbol{u}}_K \cdot \bnabla\, \overline{\rho\boldsymbol{u}\boldsymbol{u}}_K 
        + \overline{\rho\boldsymbol{u}}_K \cdot \bnabla\, \overline{\boldsymbol{u}\boldsymbol{u}}_K
    \right\rangle,\\
    \mathcal{F}^c_K &= \frac{1}{2} \left\langle  
        \overline{\boldsymbol{u}}_K \cdot \overline{\boldsymbol{F}^c}_K 
        + \overline{\rho\boldsymbol{u}}_K \cdot \overline{\left(\frac{\boldsymbol{F}^c}{\rho}\right)}_K
    \right\rangle,\\
    \mathcal{D}_K &= \frac{\mu}{2} \left\langle  
        \overline{\boldsymbol{u}}_K \cdot \nabla^2 \overline{\boldsymbol{u}}_K 
        + \overline{\rho\boldsymbol{u}}_K \cdot \overline{\left(\frac{\nabla^2 \boldsymbol{u}}{\rho}\right)}_K
    \right\rangle,\\
    \mathcal{P}_K &= -\frac{1}{2} \left\langle  
        \overline{\rho\boldsymbol{u}}_K \cdot \overline{\left(\frac{\bnabla p}{\rho}\right)}_K
    \right\rangle.
\end{aligned}
\end{equation}
The superscript $c$ above stands for $\sigma$ or $g$.
All the quantities above are cumulative functions of the filtering wavenumber $K$. Therefore, energy is injected (dissipated) where their derivatives with respect to $K$ are positive (negative).

The inter-scale energy transfer terms (advective nonlinearity, surface tension and pressure) couple different length-scales, and are non-monotonic functions of the filtering wavenumber $K$.
Further, the above equation reduces to the time evolution equation for the unfiltered kinetic energy $\mathcal{K}= 1/2\,\langle \rho u^2\rangle$  in the limit $K\to \infty$.
Note that the contributions due to the advective nonlinearity $\mathcal{N}_K$ and pressure $\mathcal{P}_K$ vanish in this limit.

\subsection{Equivalent KHM Relation using velocity-momentum correlator {\tt C1}}
We now outline a K\'arm\'an-Howarth-Monin (KHM) relation for the two-point momentum velocity correlator. 
We start by defining the average velocity-momentum correlation function for an increment vector $\br$,
\begin{equation}
    R(\boldsymbol{r}) = \frac{1}{2} \langle \bu(\boldsymbol{x}+\boldsymbol{r}) \cdot \rho(\boldsymbol{x})\bu(\boldsymbol{x})\rangle.
\end{equation}
The scale-dependent kinetic energy is then defined as the symmetrized version of the above correlator,
\begin{align}
    \mR(\br) = \frac{1}{2}\left[{R(\br) + R(-\br)}\right]=\frac{1}{4} \langle \rho\bu\cdot\bu' + \rho'\bu'\cdot\bu \rangle,
    \label{eq:khme}
\end{align}
where the superscript $(\cdot)^\prime$ indicates the value of function at position $\boldsymbol{x}^\prime$
\footnote{Note that  $R(-\br) = \avg{ \bu(\bx-\br) \cdot \rho(\bx)\bu(\bx)}/2
    = \avg{ \bu(\bx)\cdot \rho(\bx+\br)\bu(\bx+\br)}/2$.
For uniform density flows $R(\boldsymbol{r})=R(-{\boldsymbol{r}})$.}.
Starting from the Navier-Stokes equations \eqref{eq:ns} and taking a dot product with the velocity field $\bu^\prime$, we get 
\begin{equation}\label{eq:khm_crude}
 \boldsymbol{u}'\cdot \partial_t \rho \bu +\boldsymbol{u}'\cdot \bnabla\cdot \rho \bu\bu= -\boldsymbol{u}'\cdot\bnabla p + \boldsymbol{u}'\cdot\mu \nabla^{2} \bu + \boldsymbol{u}'\cdot\boldsymbol{F}^g + \boldsymbol{u}'\cdot\boldsymbol{F}^\sigma.
\end{equation}
Likewise, we write equations for 
$\rho\boldsymbol{u}\cdot\partial_t\boldsymbol{u}^\prime$, $\boldsymbol{u}\cdot\partial_t\rho'\boldsymbol{u}'$ and
$\rho'\boldsymbol{u}'\cdot\partial_t\boldsymbol{u}$. Using the above equations and after averaging over the entire domain, we get the KHM relation (see Appendix \ref{app:deriv_khm} for details):
\begin{equation}\label{eq:khm}
    \partial_t \mathcal{R}(\boldsymbol{r}) = \mathcal{N}(\boldsymbol{r}) + \mathcal{D}(\boldsymbol{r}) + \mathcal{P}(\boldsymbol{r}) + \mathcal{F}^g (\boldsymbol{r})+ \mathcal{F}^\sigma(\boldsymbol{r}).
\end{equation}
The expressions for the contributions due to advective nonlinearity $\mathcal{N}(\br)$, viscous dissipation $\mathcal{D}(\br)$, pressure $\mathcal{P}(\br)$, surface tension $\mathcal{F}^\sigma(\br)$, and buoyancy injection $\mathcal{F}^g(\br)$ are: 
\begin{equation}\label{eq:khm_contri}
\begin{aligned}
    \mathcal{N}(\boldsymbol{r}) &= \frac{1}{4}\bnabla_{\br} \cdot \langle \delta(\rho\bu)\cdot\delta\bu\,\delta\bu\rangle,\\
    \mathcal{D}(\boldsymbol{r}) &= \frac{\mu}{4}\left \langle \bu' \cdot \nabla^2 \bu + \rho \bu \cdot \frac{\nabla'^2\bu'}{\rho'} + \bu \cdot \nabla'^2\bu' + \rho' \bu' \cdot \frac{\nabla^2\bu}{\rho} \right \rangle,\\ 
        \mathcal{P}(\boldsymbol{r}) &= -\frac{1}{4} \left \langle \rho \bu \cdot \frac{\bnabla'p'}{\rho'}  + \rho' \bu' \cdot \frac{\bnabla p}{\rho} \right \rangle, \,\text{and} \\
    \mathcal{F}^c(\boldsymbol{r})&= \frac{1}{4}\left \langle \bu' \cdot \boldsymbol{F}\,^c + \rho \bu \cdot \frac{\boldsymbol{F}^{c}\,'}{\rho'} + \bu \cdot \boldsymbol{F}^c\,' + \rho' \bu' \cdot \frac{\boldsymbol{F}^c}{\rho} \right \rangle,
\end{aligned}
\end{equation}
where $c$ stands for either $g$ (buoyancy) or $\sigma$ (surface tension). We would like to highlight the similarity in the structure of the budget contributions in \eqref{eq:khm_contri} and \eqref{eq:f1_contri}.
The above correlation functions can be exactly mapped to the quantities in the filtered budget in \eqref{eq:f1_contri} (obtained by sharp-spectral filters) using equation \eqref{eq:WienerKhnichin}.

The KHM relations derived for compressible flows \citep{GaltierBanerjee2011,Lai_Charonko_Prestridge_2018} are identical to \eqref{eq:khm} in the incompressible limit ($\bnabla \cdot \bu=0$). 
Note that \eqref{eq:khm} is identical to the KHM relation derived in the recent work of \citet{Fabien_2025}.
Furthermore, for iso-density flows, \eqref{eq:khm} reduces to the standard KHM relation as in \citet{Frisch_1995}.
To make comparison with the scale-by-scale budget equation, we perform an angular averaging and obtain the contribution from the isotropic sector for different terms.
\subsection{Scale by scale budget for Favre filtered energy $\mathcal{E}_K=(1/2)\langle \overline{\rho \bu}_K \cdot \overline{\rho \bu}_K/\overline{\rho}_K \rangle$ ({\tt F2})}\label{subsec:favre}
The following discussion closely follows \citet{PandeyMitraPerlekar2023} and the supplemental material within.
For a field $\psi$, the corresponding Favre filtered field is defined as 
\begin{align}
\widetilde{\psi}_K= \frac{\overline{\rho \psi}_K}{\overline{\rho}_K}.
\end{align}
From the Navier-Stokes equations, we can obtain the following evolution equations for the filtered density, and the filtered momentum field:
\begin{eqnarray}
\partial_t \overline{\rho}_K + \nabla \cdot \overline{\rho}_K \widetilde{\bu}_K &=&0,~\rm{and}\\
\partial_t \overline{\rho}_K \widetilde{\bu}_K + \bnabla\cdot\overline{\rho\bu\bu}_K &=& -\bnabla\,\overline{p}_K +\mu\nabla^2\overline{\bu}_K + \overline{\boldsymbol{F}^g}_K+\overline{\boldsymbol{F}^\sigma}_K.
\label{eq:filF2}
\end{eqnarray}
As evident from the density equation above, Favre velocity is the advection velocity for the filtered density.
From the above equations, we obtain the following scale-by-scale energy budget equation:

\begin{equation}
    \partial_t \mathcal{E}_K = \mathcal{N}_K + \mathcal{D}_K + \mathcal{P}_K + \mathcal{F}^g_K + \mathcal{F}^\sigma_K, 
\end{equation}
with 
\begin{eqnarray}\label{eq:f2_contri}
   \nonumber \mathcal{E}_K &=& \langle \overline{\rho}_K |\tilde{\bu}_K|^2 \rangle/2, \\
    \nonumber\mathcal{N}_K &=&  \left\langle\overline{\rho}_K \bnabla\,  \overline{\boldsymbol{u}}_K: ( \widetilde{\boldsymbol{u}\boldsymbol{u}}_K -\widetilde{\boldsymbol{u}}_K\,\widetilde{\boldsymbol{u}}_K)\right\rangle, \\ 
    \nonumber\mathcal{D}_K &=&  \left\langle\mu  \widetilde{\boldsymbol{u}}_K\cdot \nabla^2 \overline{\boldsymbol{u}}_K \right\rangle,  \\
    \nonumber\mathcal{P}_K &=& - \left\langle  \widetilde{\boldsymbol{u}}_K\cdot \bnabla \,\overline{p}_K \right\rangle,~\rm{and} \\
 \mathcal{F}^c_K &=&  \left\langle  \widetilde{\boldsymbol{u}}_K\cdot\overline{\boldsymbol{F}^c}_K \right\rangle,
\end{eqnarray}
with $c$ again standing in for either buoyancy ($g$) or surface tension ($\sigma$).
There is no pressure dilation contribution ($\sim \overline{p}_K\bnabla\cdot\overline{\bu}_K$) since the filtered velocity field is incompressible, and therefore the pressure contribution is entirely baropycnal \citep{Aluie_2011,Aluie2013,LeesAluie_2019}.\newline 
It was shown in \citet{Aluie2013} that the injection, here $\mathcal{F}^g_K$, is localized to large-scales for large-scale stirring.
\citet{Aluie2013} also mathematically showed that the viscous term in the Favre scheme $\mathcal{D}_K$ has negligible effects in the inertial-range in the infinite Reynolds number limit.
These statements, in general, are not true for the other definitions as demonstrated in \citet{ZhaoAluie_2018}.

\section{Results \label{sec:Results}}
Using runs {\tt R1} and {\tt R2}, we present below a comparative study of the scale-by-scale energy budget obtained using definitions {\tt F1}, {\tt F2} and {\tt C1}. 
We perform this analysis in the statistically steady-state, where $\partial_t {\mathcal E}_K=0$ and $\partial_t {\mathcal R}(\br)=0$.
In what follows, angular brackets denote averaging over space and fifteen independent realizations of the flow field.

\subsection{Equivalence of the filtered budget {\tt F1} and the KHM relation {\tt C1}}
We first verify the equivalence of the velocity-momentum filtered budget {\tt F1} and the corresponding KHM relation with correlator {\tt C1}.

\subsubsection{Small Atwood Number $\At=0.04$ (Run {\tt R1})}
In the small Atwood number limit ($\rho_B/\rho_L\approx 1$ and $\rho_a\approx \rho_L$), we can use the Boussinesq approximation, wherein the density variations are retained only in the buoyancy term $\boldsymbol{F}^g\approx (\rho_B-\rho_L) \phi {\boldsymbol{g}}$ and the density is taken to be a constant everywhere else.
With this approximation, equations~\eqref{eq:ns} simplify to \citep{Pandey_Mitra_Perlekar_2022}:
\begin{subequations}
\begin{align}
\rho_a \text{D}_t \bu &= -\bnabla p + \mu \nabla^{2} \bu + (\rho_B-\rho_L) \phi {\boldsymbol{g}} + \boldsymbol{F}^\sigma, \\
 \text{D}_t \phi &= 0,~{\rm and}~ \bnabla   \cdot \bu  = 0.
\end{align}
\label{eq:nsb}
\end{subequations}
The scale-by-scale budget is obtained by replacing $\rho$ with $\rho_a$ in the definitions {\tt F1-F4} detailed in section~\ref{sec:ScaleByScale}. 
It is evident that the resulting equation for scale-by-scale transfer is identical for all the cases:

\begin{equation}
    \partial_t \mathcal{E}_K = \mathcal{N}_K + \mathcal{D}_K + \mathcal{P}_K + \mathcal{F}^g_K + \mathcal{F}^\sigma_K, \label{eq:fb}
\end{equation}
with 
\begin{equation}
\begin{aligned}
\mathcal{E}_K &= \rho_a \langle |\overline{\bu}_K|^2 \rangle/2, \\
\mathcal{N}_K &=  \rho_a \left\langle \bnabla\,  \overline{\boldsymbol{u}}_K: ( \overline{\boldsymbol{u}\boldsymbol{u}}_K -\overline{\boldsymbol{u}}_K\,\overline{\boldsymbol{u}}_K)\right\rangle, \\ 
    \mathcal{D}_K &=  -\mu \left\langle \nabla \overline{\boldsymbol{u}}_K|^2 \right\rangle,~\rm{and} \\
 \mathcal{F}^c_K &=  \left\langle  \overline{\boldsymbol{u}}_K\cdot\overline{\boldsymbol{F}^c}_K \right\rangle,
\end{aligned}
\end{equation}
with $c$ again standing for either buoyancy ($g$) or surface tension ($\sigma$).

Due to the incompressibility constraint ($\bnabla\cdot\bu=0$ and $\bnabla\cdot\overline{\bu}_K=0$), the pressure term $\mathcal{P}_K = -\langle \overline{\bu}_K\cdot\bnabla\, \overline{p}_K \rangle = \langle (\bnabla\cdot\overline{\bu}_K) \overline{p}_K \rangle = 0$ and thus
does not contribute to the energy transfers across scales \citep{Frisch_1995}.
Similarly, replacing $\rho$ with $\rho_a$ in \eqref{eq:khm_contri} gives the  KHM relation corresponding to \eqref{eq:fb}. 
Using the incompressibility condition, the pressure contribution ${\mathcal P}(r)$ vanishes also for eq. \eqref{eq:khm_contri}.

We now compare the energy budget obtained using the scale-by-scale analysis \eqref{eq:fb} and the KHM relation \eqref{eq:khm}. To make the comparison clear, we have used the empirical correspondence $K\cdot r \sim \sqrt{3}$ suggested by \citet{Hellinger_2021b} in the KHM relation \eqref{eq:khm}. In addition, we plot all the quantities as a function of $K/K_D$, where $K_D=2\upi/D$ is the wavenumber corresponding to the diameter of the bubble.

The budget contributions obtained using the two methods show excellent agreement. The buoyancy term ${\mathcal F}^g_K$ performs a net injection at scales comparable to the bubble diameter, while the viscous term ${\mathcal D_K}$ dissipates energy at small scales.
The energy is transferred from the large scales to small scales through both surface tension ${\mathcal F}^\sigma_K$ and nonlinear advection ${\mathcal N}_K$ \citep{Pandey_Ramadugu_Perlekar_2020,PandeyMitraPerlekar2023,RAMIREZ2024104860}.  
Note that the mechanism for energy transfer due to surface tension is now well-established for both bubbly flows \citep{Pandey_Perlekar_Mitra2019,Pandey_Mitra_Perlekar_2022,Fabien_2025} and emulsions \citep{Dodd_Ferrante_2016,Perlekar_2019,Brandt_2022}.
In particular, bubbles absorb energy due to the buoyancy injection as they stretch and deform, this energy is released at small scales as they relax.

\begin{figure}
\centering
  \includegraphics[width=0.485\linewidth]{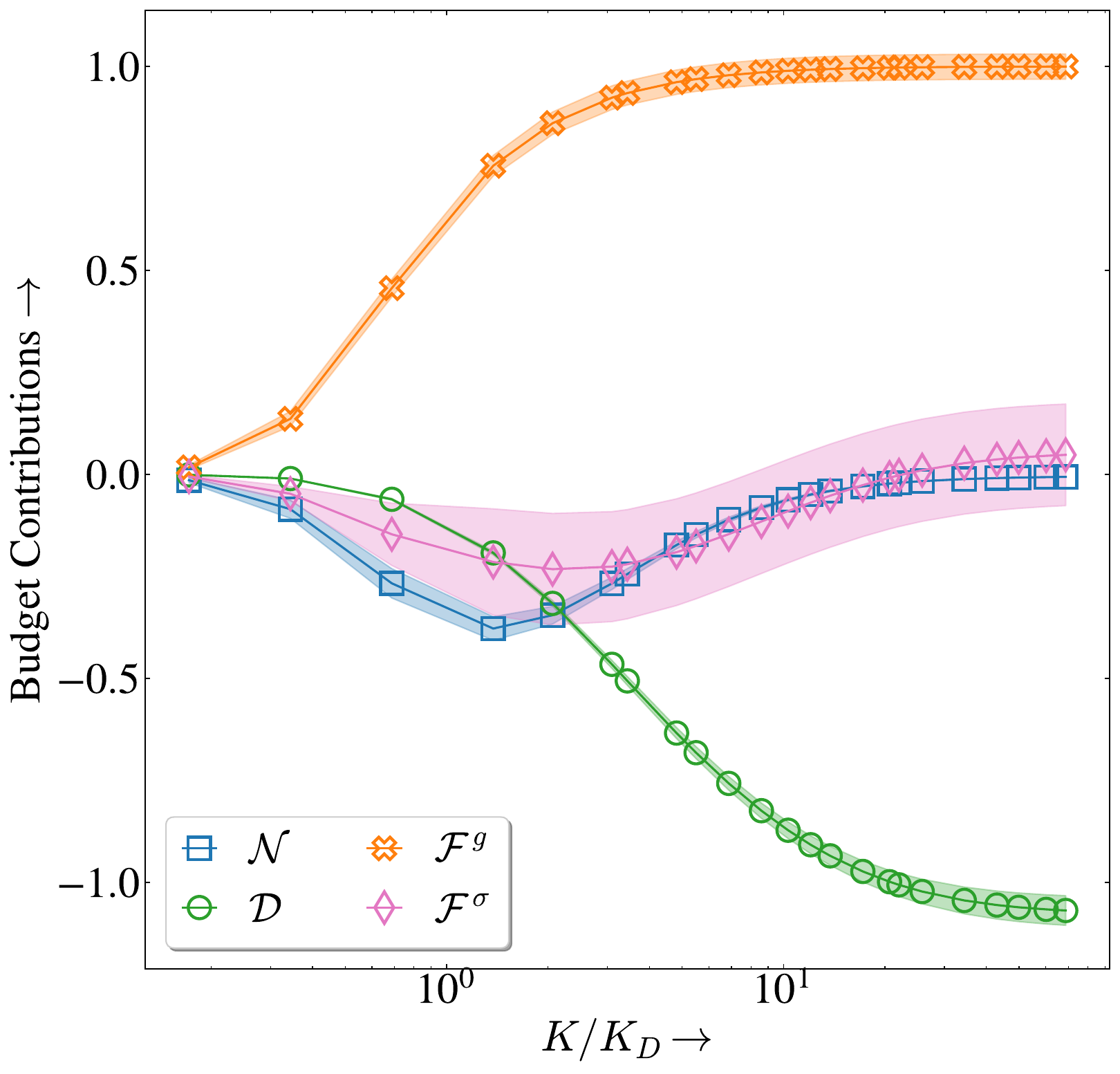}
  \includegraphics[width=0.485\linewidth]{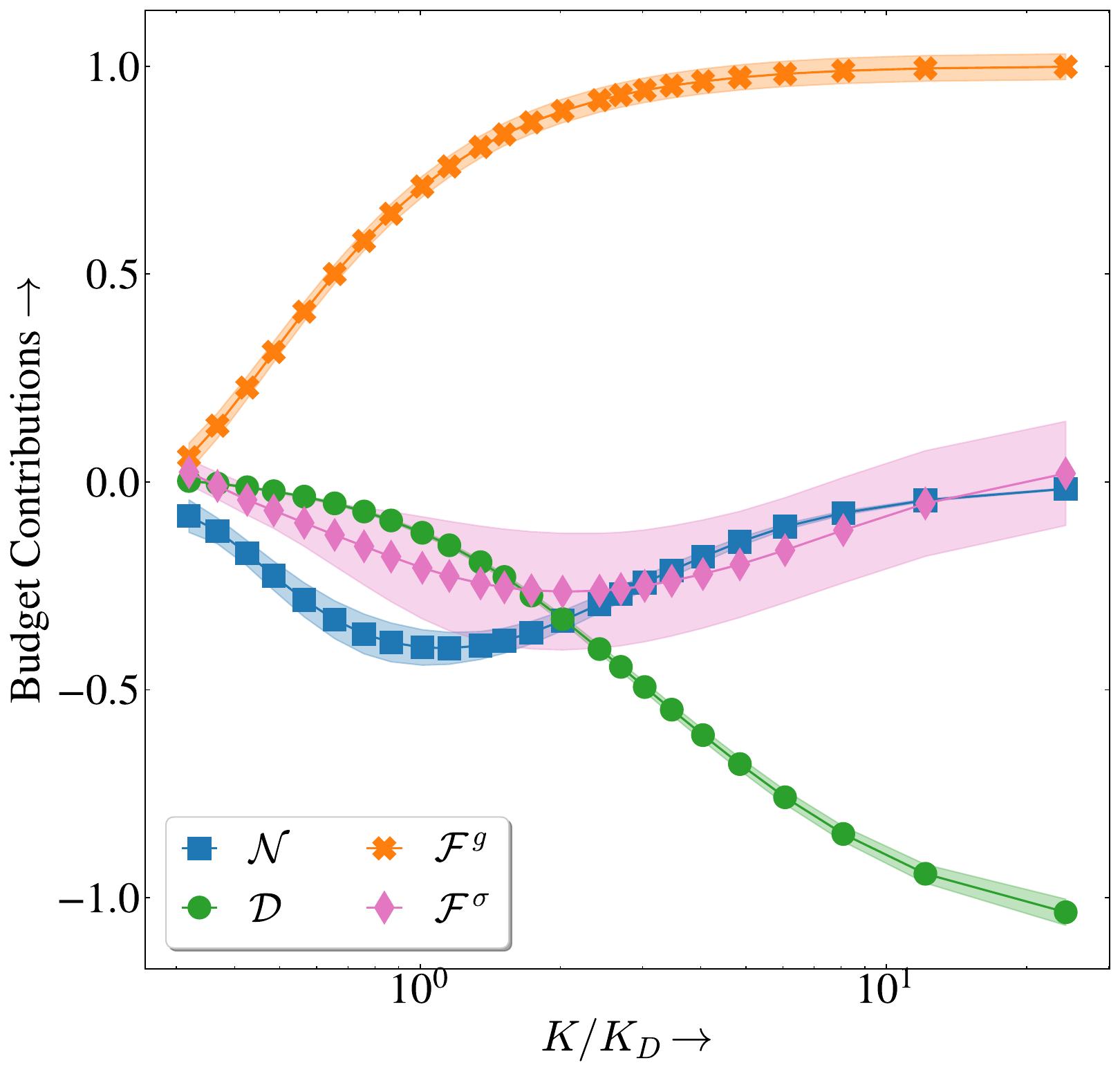}
  
\caption{Scale-by-scale budget using {\tt F1} filtering \eqref{eq:f1_contri}  (left) and the KHM relation {\tt C1} \eqref{eq:khm_contri} (right, with $K=\sqrt{3}/r$) for small Atwood number (run {\tt R1}). The pressure contribution is identically zero, hence not shown here. The shading indicates one standard deviation variations. In both the plots, all the budget contributions (vertical axis) have been normalized by the mean energy  injection rate $\epsilon^g\equiv \langle {\bu} \cdot {\boldsymbol{F}^g} \rangle$.\label{fig:hga_r4}}
\end{figure}

\subsubsection{Large Atwood Number $\At=0.8$ (Run {\tt R2})}
\begin{figure}
\centering
  \includegraphics[width=0.485\linewidth]{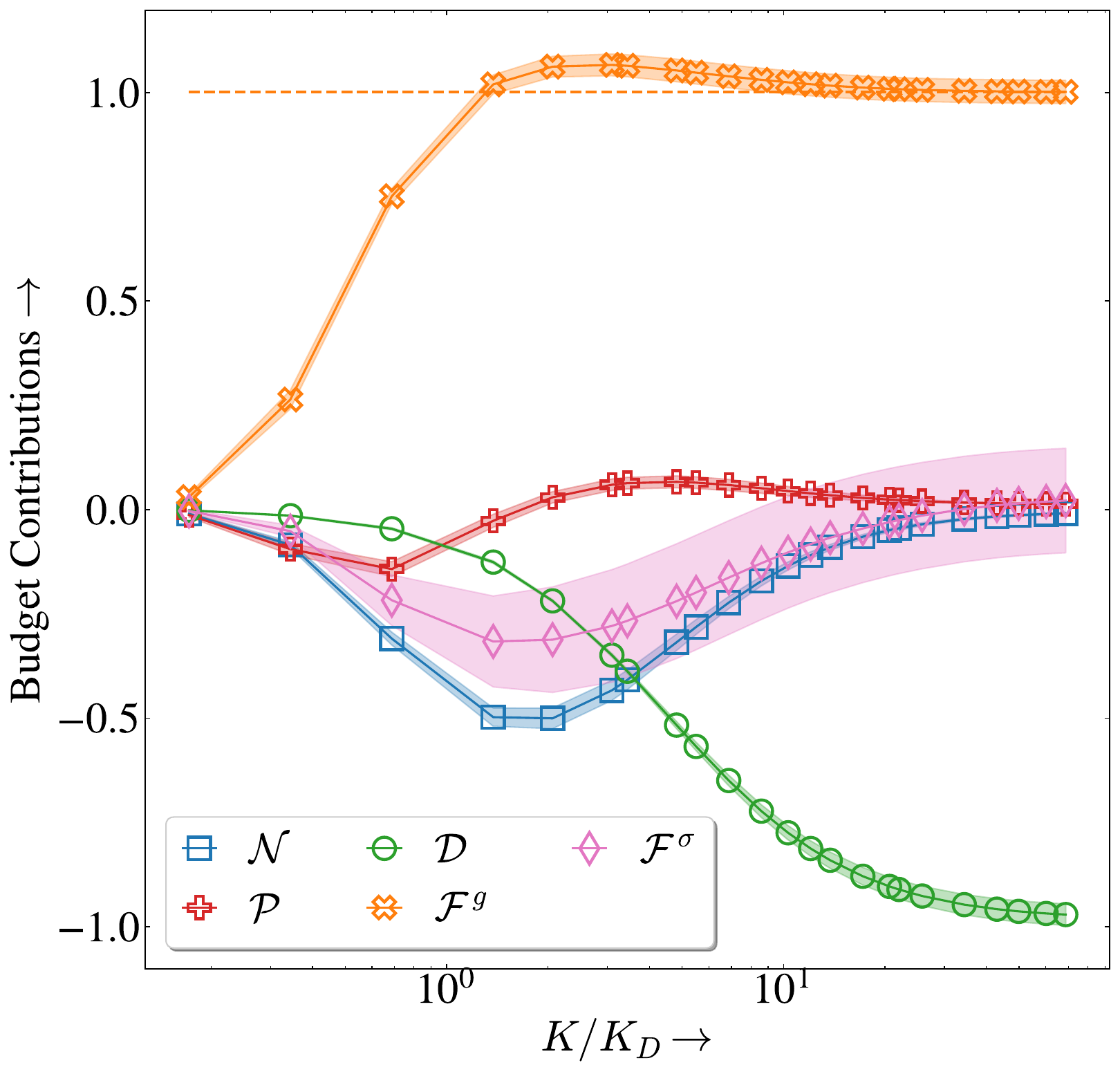}
  \includegraphics[width=0.485\linewidth]{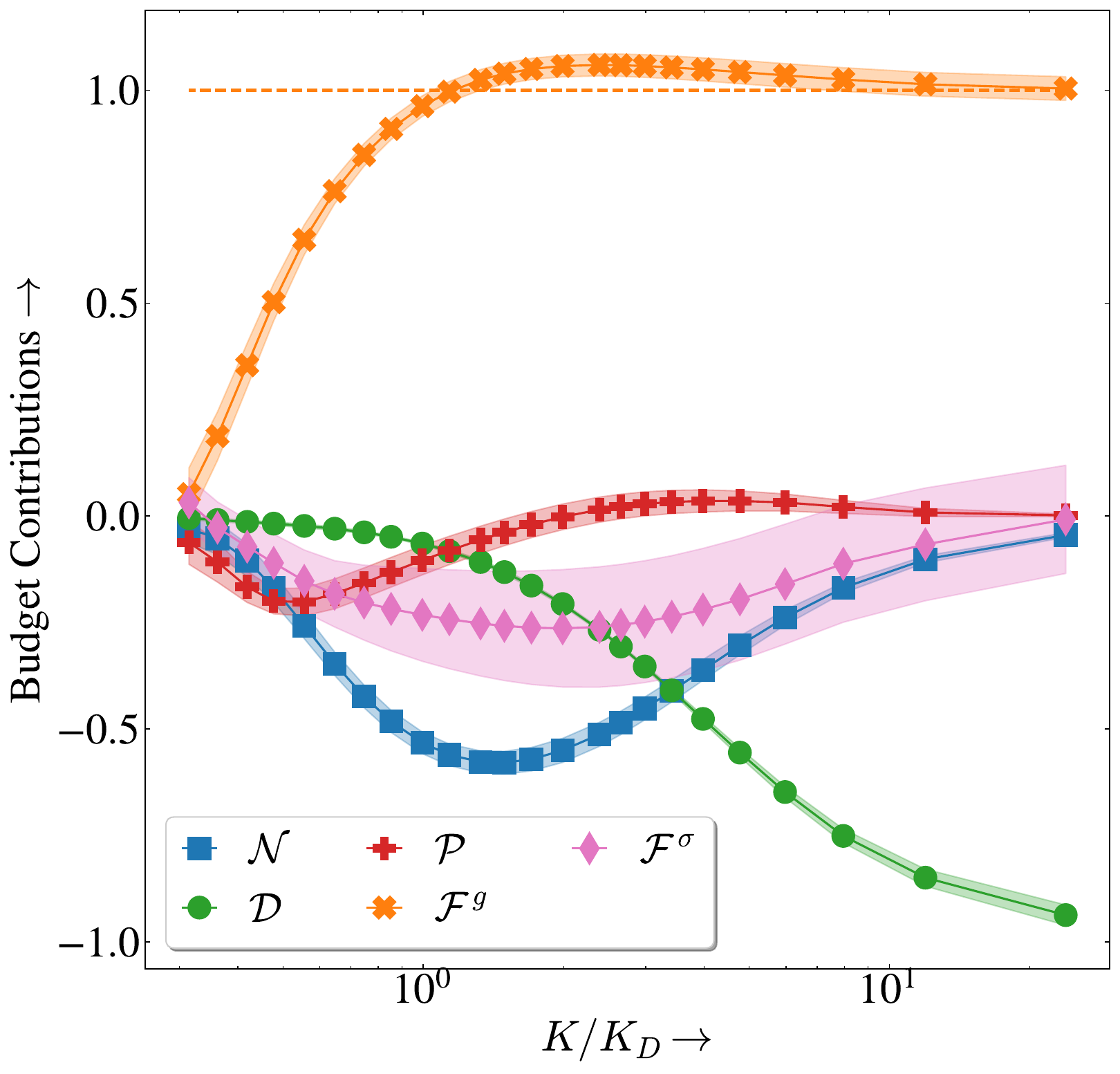}
\caption{Scale-by-scale budget using {\tt F1} filtering \eqref{eq:f1_contri}  (left) and the KHM relation {\tt C1} \eqref{eq:khm_contri} (right, with $K=\sqrt{3}/r$) for large Atwood number (run {\tt R2)}. The shading indicates one standard deviation variations. The orange dashed line indicates the unfiltered buoyancy injection to convey the non-monotonic behavior of the buoyant flux. In both the plots, all the budget contributions (vertical axis) have been normalized by the mean energy  injection rate $\epsilon^g\equiv \langle {\bu} \cdot {\boldsymbol{F}^g} \rangle$.
\label{fig:hga_r8_nf}}
\end{figure}

Comparison between different contributions to the filtered budget {\tt F1}  \eqref{eq:f1_contri} and the KHM relation {\tt C1} \eqref{eq:khm_contri} for run {\tt R2} is shown in figure \ref{fig:hga_r8_nf}. 
The excellent agreement between different contributions for the two methods is again evident. 
This is consistent with previous studies on compressible turbulence \citep{Hellinger_2021a}.
The dependence of nonlinear advection ${\mathcal N}_K$, surface tension ${\mathcal F}^\sigma_K$, and viscous dissipation ${\mathcal D}_K$ on the filtering wave number $K$ is similar to the small $\At$ case discussed in the previous section. Their interpretation in the scale-by-scale budget is also identical.

However, the buoyancy contribution for large $\At$ varies non-monotonically with the filtering wavenumber $K$ (see ${\mathcal F}^g_K$ in figure~\ref{fig:hga_r8_nf}). It initially rises, reaching a peak around $K\sim K_D$, then declines and saturates to $\epsilon^g$.
This suggests that, in addition to injection, buoyancy plays a role in the inter-scale energy transfers. 
This assertion is further justified by noting that the contribution of the buoyancy term in the time evolution equation of the residual energy (at scales $>K$), $\mathcal{K}-\mathcal{E}_K$, is equal to ${\mathcal F}_K^{g,\,{\rm small}}=\epsilon^g-\mathcal{F}^g_K$. From the plot of $\mathcal{F}^g_K$, it is evident that ${\mathcal F}_K^{g,\,{\rm small}}$ remains positive for $K<K_D$ and turns negative when $K>K_D$.
This sign reversal between the resolved and residual buoyancy contribution indicates an inverse energy transfer from small scales ($K>K_D$) to intermediate scales ($K\approx K_D$) \citep{Pope_2000}. 
Note that similar non-monotonic behavior of the energy injection is also observed in definitions {\tt F3} and {\tt F4} (see figure \ref{fig:density_buoyancy_compare} (b) in Appendix \ref{app:all_buoyancy}).

Furthermore, the pressure contribution $\mathcal{P}_K$ is no longer negligible, and it transfers energy from both the small and large scales (where ${\rm d}\mathcal{P}_K/{\rm d}K<0$) to the intermediate scales (where ${\rm d}\mathcal{P}_K/{\rm d}K>0$).

How can an injection mechanism do transfer across scales and how can pressure simultaneously do a transfer to both large and small scales? 
We do not address this immediately, but rather first investigate the Favre filtered {\tt F2} energy budget in the next section.

\subsection{Favre filtered budget {\tt F2} for the large Atwood number case (Run {\tt R2})}
\begin{figure}
\centering
  \centering
  \includegraphics[width=0.92\linewidth]{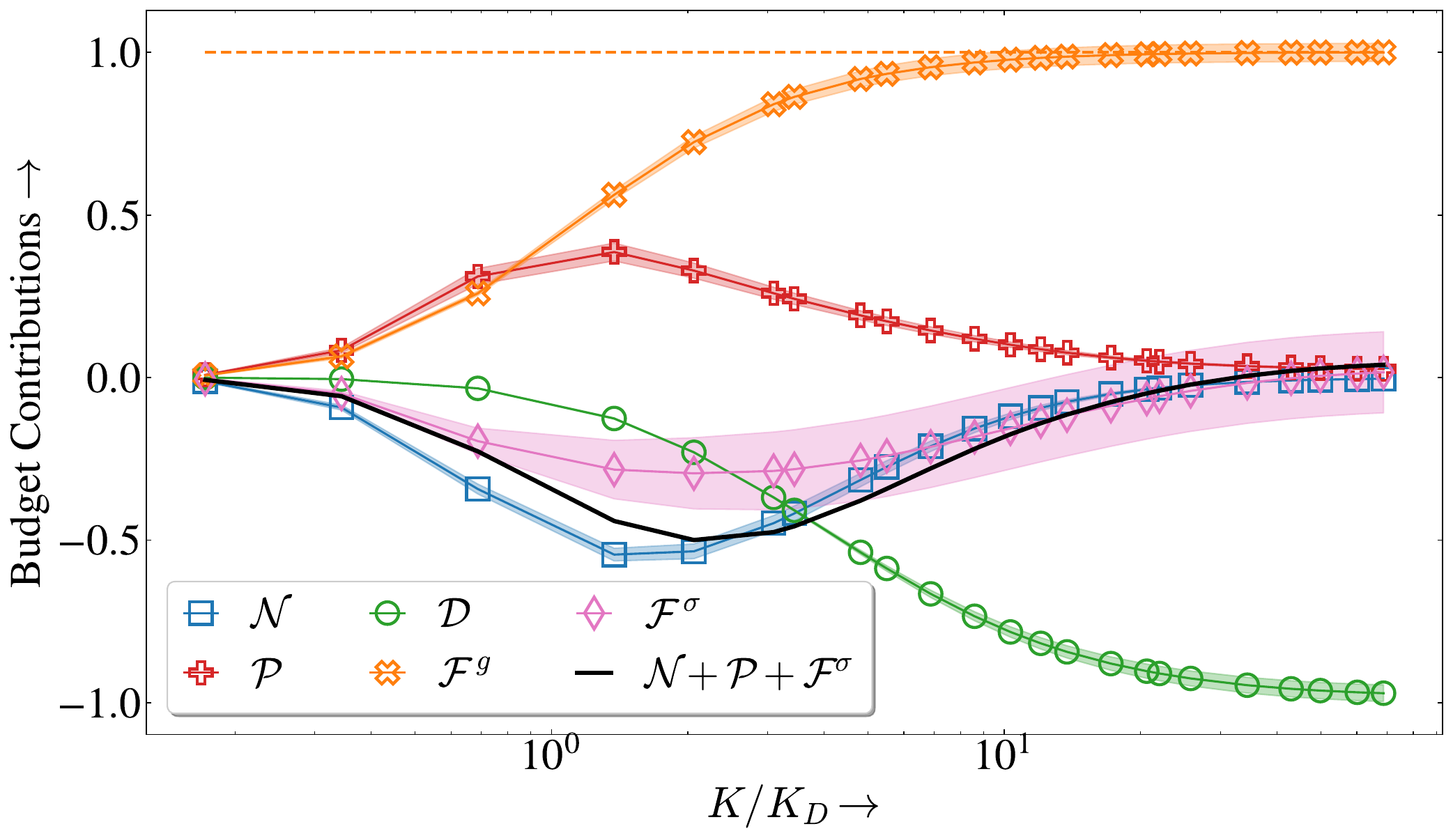}
\caption{Scale-by-scale budget using the Favre filtering ({\tt F2}) \eqref{eq:f2_contri} for the large Atwood number (run {\tt R2}). The shading indicates one standard deviation variations. All the budget contributions (vertical axis) have been normalized by the mean energy  injection rate $\epsilon^g\equiv \langle {\bu} \cdot {\boldsymbol{F}^g} \rangle$. In contrast with the {\tt F1} budget (figure \ref{fig:hga_r8_nf}), the buoyancy term increases monotonically.}
\label{fig:hga_r8_fav}
\end{figure}

The contribution of the different terms in the Favre-filtered budget \eqref{eq:f2_contri} for the large Atwood number case is shown in figure \ref{fig:hga_r8_fav}. A comparison with figure \ref{fig:hga_r8_nf} reveals that the contributions of nonlinear advection $\mathcal{N}_K$, viscous dissipation $\mathcal{D}_K$, and surface tension $\mathcal{F}^\sigma_K$ are qualitatively identical to the scale-by-scale budget obtained for {\tt F1} \eqref{eq:f1_contri}.
However, the buoyancy and pressure contributions are very different.
The cumulative buoyancy injection is a monotonically increasing function and saturates at large $K$ to $\epsilon^g$ \citep{Aluie2013}. 
Thus, in contrast to the ${\tt F1}$ definition, buoyancy injects energy at all scales.  
Perhaps the most striking is the pressure contribution, which for the Favre filtered budget performs an inverse transfer from small scales (where ${\rm d}\mathcal{P}_K/{\rm d}K<0$) to large scales (where ${\rm d}\mathcal{P}_K/{\rm d}K>0$). This inverse baropycnal transfer \citep{Aluie2013} is consistent with previous studies on forced compressible turbulence \citep{Wang_2013,LeesAluie_2019}. 
However, the sum of all transfer terms (advection, pressure, and surface tension), the black solid line in figure~\ref{fig:hga_r8_nf}, still shows a net transfer of energy from large (${\rm d}\mathcal{P}_K/{\rm d}K+{\rm d}\mathcal{N}_K/{\rm d}K+{\rm d}\mathcal{F}^\sigma_K/{\rm d}K<0$) to small scales (${\rm d}\mathcal{P}_K/{\rm d}K+{\rm d}\mathcal{N}_K/{\rm d}K+{\rm d}\mathcal{F}^\sigma_K/{\rm d}K>0$).

\begin{figure}
\centering
  \includegraphics[width=0.92\linewidth]{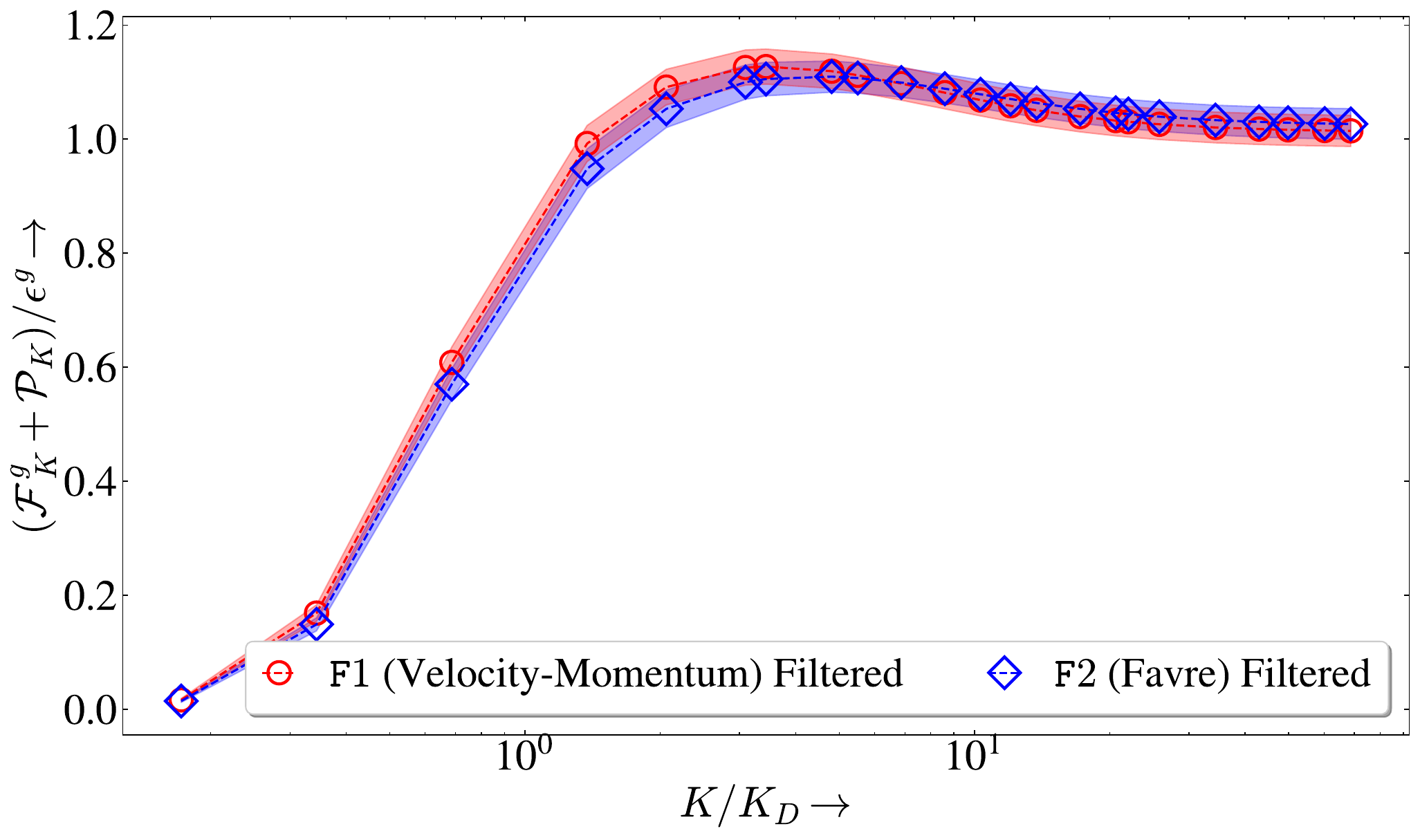}
  \caption{The sum of the baropycnal and the buoyancy contribution to the scale-by-scale budget (run {\tt R2}) for both {\tt F1} and {\tt F2} definitions.
    \label{fig:hga_r8_abaro_grav}}
\end{figure}%

\subsection{Reconciling definitions {\tt F1} and {\tt F2}}\label{sec:reconcile}
Evidently, the definitions {\tt F1} and {\tt F2} give strikingly different results for buoyancy $\mathcal{F}^g_K$ and pressure $\mathcal{P}_K$ contributions. 
Since the other contributions are similar, one finds that, not surprisingly, the sum $\mathcal{F}^g_K + \mathcal{P}_K$ is nearly the same in the two definitions as well, see figure \ref{fig:hga_r8_abaro_grav}.

In the Favre-filtered framework {\tt F2}, the buoyancy $\mathcal{F}^g_K$ injects energy and is not involved in transfers, while the pressure term
does an inverse energy transfer from small scales to large scales.
In contrast, for the definition {\tt F1}, both buoyancy and pressure contribute to the inter-scale transfers.

Motivated by this, our subsequent analysis explores modifications to {\tt F1} to ensure that buoyancy contribution only injects energy , and whether this allows us to reconcile definitions {\tt F1} and {\tt F2}.

\subsubsection{Decomposing the buoyancy contribution in {\tt F1}}
We propose to modify {\tt F1} by extracting the transfer part of $\mathcal{F}_K^g$ and adding it to the pressure term $\mathcal{P}_K$.
Recall that the buoyancy term in {\tt F1} has two contributions \eqref{eq:f1_contri},
\begin{equation}\label{eq:F1grav_break}
    \mathcal{F}\,^{{\tt F1},\,g}_K =    \frac{1}{2}\underbrace{ \left\langle \overline{\bu}_K\cdot\overline{\boldsymbol{F}^g}_K\right\rangle}_{\mathcal{I}_K} + \frac{1}{2}\underbrace{\left\langle\overline{\rho\bu}_K \cdot \overline{\left(\frac{\boldsymbol{F}^g}{\rho}\right)}_K  \right\rangle}_{\mathcal{J}_K}.
\end{equation}
The first term $\mathcal{I}_K$ is a pure injection term and is close to $\mathcal{F}^{{\tt F2},\,g}_K$ for all $K$  (see Appendix \ref{app:all_buoyancy} and \citep{Aluie2013}), unlike $\mathcal{J}_K$.
Note that the structure of the buoyancy contribution term in the Favre budget $\langle \widetilde{\bu}_K \cdot \overline{\boldsymbol{F}^g}_K \rangle\equiv\langle \overline{{\rho \bu}}_K \cdot \overline{\boldsymbol{F}^g}_K/\overline{\rho}_K\rangle$ is similar to $\mathcal{J}_K$.
Therefore, we identify the transfer part in $\mathcal{J}_K$ as $\mathcal{J}_K-\mathcal{F}_K^{\tt{F2},\,g}$,
\begin{align}\label{eq:tk_non_monoF1}
\nonumber
\mathcal{J}_K&=\left\langle\overline{\rho\bu}_K \cdot \overline{\left(\frac{\boldsymbol{F}^g}{\rho}\right)}_K  \right\rangle \\ 
&=  \left\langle\overline{\rho\bu}_K \cdot \frac{\overline{\boldsymbol{F}^g}_K}{\overline{\rho}_K}  \right\rangle + \left\langle\overline{\rho\bu}_K \cdot \left[\overline{\left(\frac{\boldsymbol{F}^g}{\rho}\right)}_K  -\frac{\overline{\boldsymbol{F}^g}_K}{\overline{\rho}_K}\right]  \right\rangle. 
\end{align} 
The above decomposition can be thought of as the contribution from the large-scale (resolved) acceleration $\overline{\boldsymbol{F}^g}_K/{\overline\rho}_K$ and a residual (sub-grid) contribution $\overline{\boldsymbol{F}^g/\rho}\,_K - \overline{\boldsymbol{F}^g}_K/{\overline\rho}_K$ which is involved in transfers across scales.

\begin{figure}
\centering
  \centering
  \includegraphics[width=0.92\linewidth]{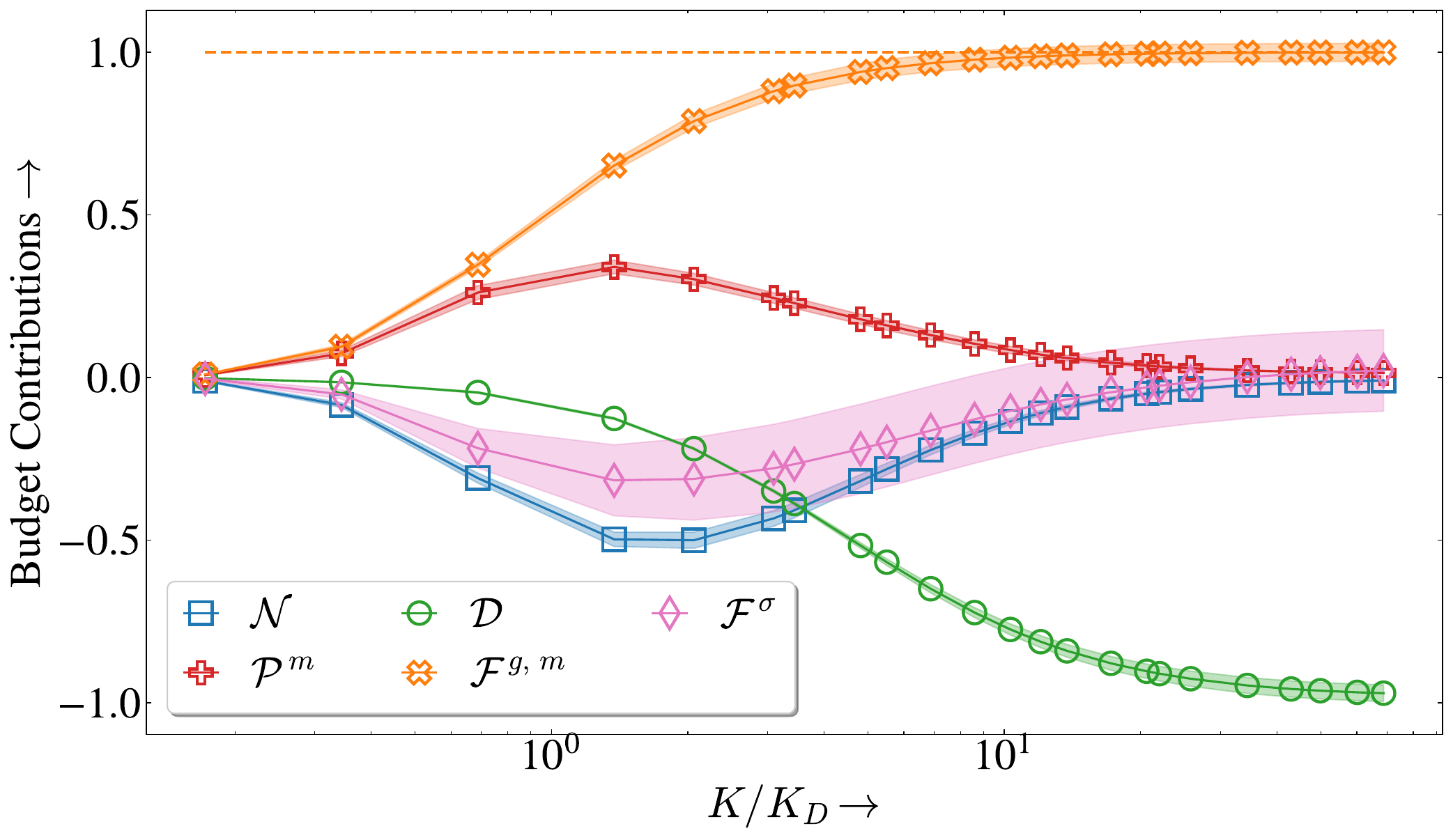}
  \label{fig:hga_r8_afav}
\caption{{\tt F1} filtered scale-by-scale budget for the large Atwood number run {\tt R2} with the modified pressure and buoyancy contributions (as in \eqref{eq:f1_corrected}). The shading indicates one standard deviation variations. All the budget contributions (vertical axis) have been normalized with the mean injection. 
Notice the monotonic behavior of the buoyancy term, contrast with figure \ref{fig:hga_r8_nf}. Further, the pressure term $\mathcal{P}_K$ now transfers energy definitively from small to large scales.}
\label{fig:hga_r8_F1corrected}
\end{figure}

Indeed, if we add the residual contribution in buoyancy to the pressure transfer, we get the following modified equation for the {\tt F1} budget,
\begin{equation}
    \partial_t \mathcal{E}_K = \mathcal{N}_K + \mathcal{D}_K + \mathcal{P}\,^{m}_K + \mathcal{F}^{g,\,m}_K + \mathcal{F}^\sigma_K,
\end{equation}
with the following definitions of the modified buoyancy and pressure terms,
\begin{eqnarray}\label{eq:f1_corrected}
 \mathcal{F}^{g,\,m}_K &=& \frac{1}{2} \left\langle  \overline{\boldsymbol{u}}_K\cdot \overline{\boldsymbol{F}^g}_K + \overline{\rho\bu}_K \cdot \frac{\overline{\boldsymbol{F}^g}_K}{\overline{\rho}_K}\right\rangle,~{\rm and} \\
\mathcal{P}^m_K &=& -\frac{1}{2} \left\langle \overline{\rho\boldsymbol{u}}_K\cdot\overline{\left(\frac{\bnabla p}{\rho}\right)}_K\right\rangle + \left\langle\overline{\rho\bu}_K \cdot \left[\overline{\left(\frac{\boldsymbol{F}^g}{\rho}\right)}_K  -\frac{\overline{\boldsymbol{F}^g}_K}{\overline{\rho}_K}\right]  \right\rangle.    
\end{eqnarray}
The superscript $m$ indicates modified quantities.
With these new definitions, in figure \ref{fig:hga_r8_F1corrected}, we plot the modified budget.
Now all the transfers are consistent with the Favre-filtered framework. 
In particular, the injection term monotonically increases and the transfer due to modified pressure is indeed from small-scales to large-scales.

Thus, on appropriately identifying the residual transfer terms and the pure injection contributions, we find that both the definitions of evaluating the budget give similar results. 

\section{Spatial Distribution of the Buoyancy Contribution in {\tt F1} and {\tt F2}}
\begin{figure}
\centering
  \includegraphics[width=0.42\linewidth]{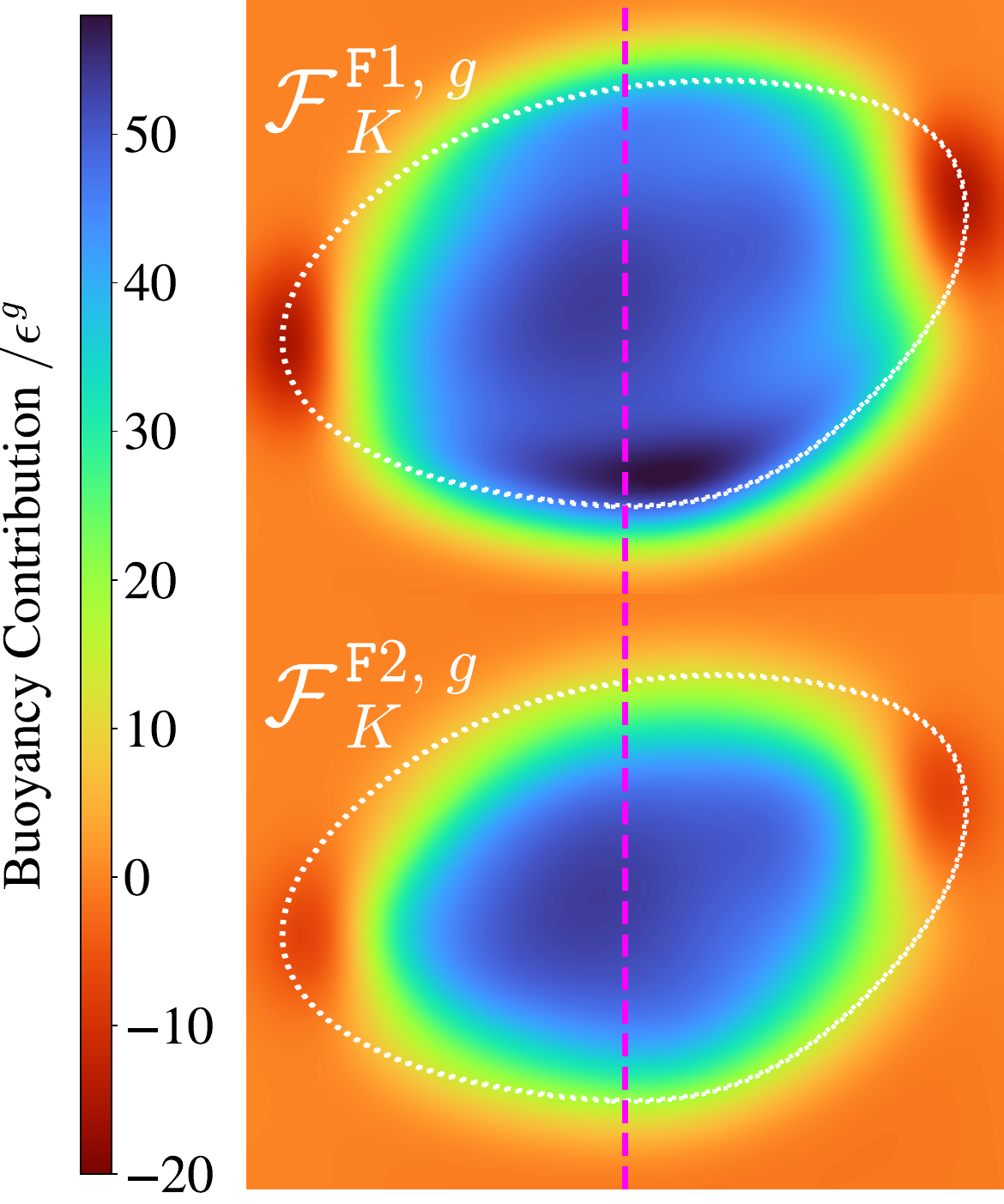}\quad 
  \includegraphics[width=0.48\linewidth]{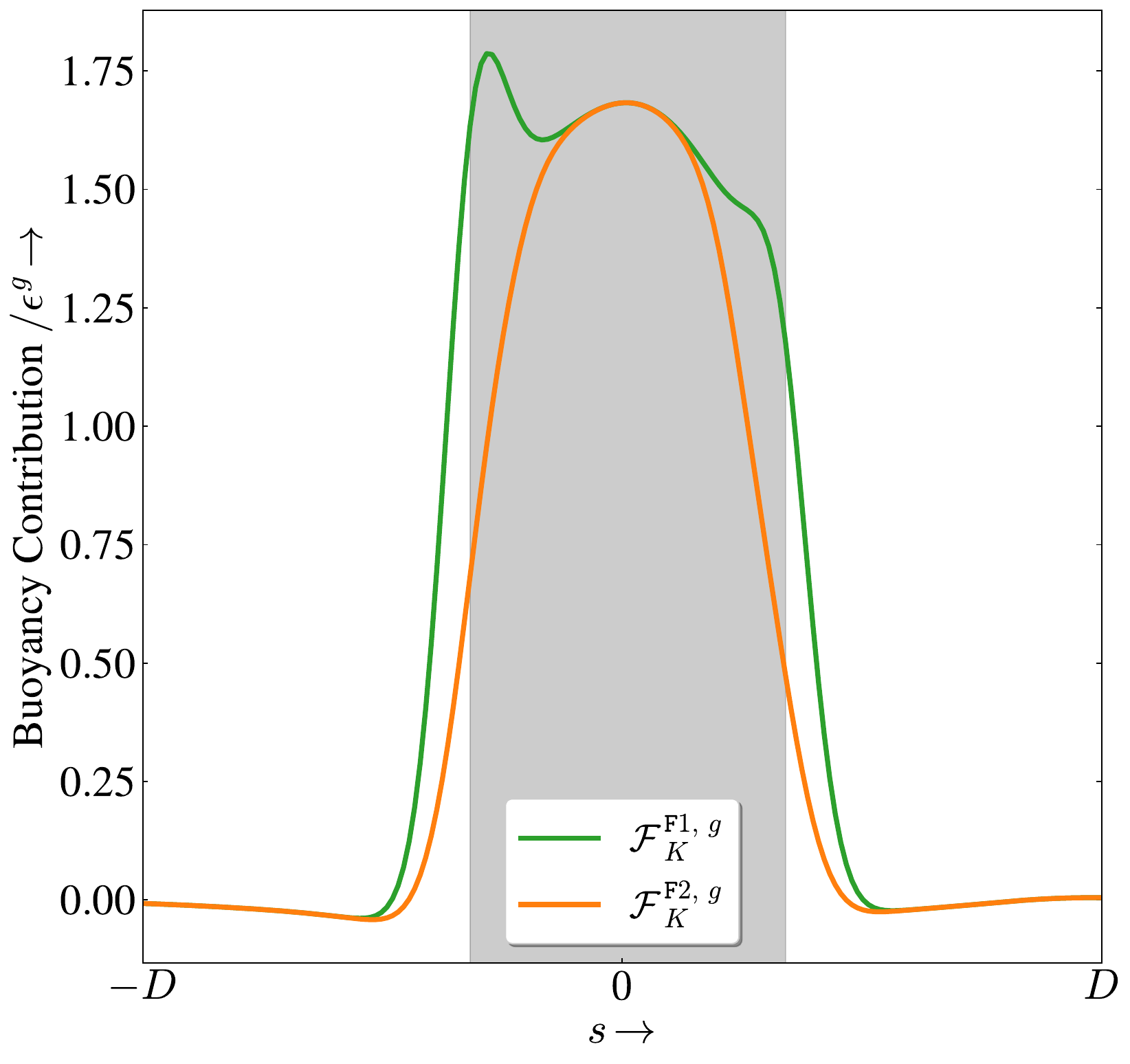}
\caption{2$d$ slice of of the buoyancy contribution for {\tt F1} \eqref{eq:f1_contri} and {\tt F2} \eqref{eq:f2_contri} (left, for the top right bubble in figure \ref{fig:hga_r8_representative} (c)) and corresponding 1$d$ vertical cut through the center, marked by the dashed line, (as a function of the distance from bubble center $s$) (right). The filtering wavenumber $K\approx2K_D$.
\label{fig:injectioncompare}}
\end{figure}

To investigate the physical origin of the discrepancy between the buoyancy injection mechanisms in the definitions {\tt F1} and {\tt F2}, we now contrast the filtered injection fields:
\begin{align}
 {\mathcal F}^{{\tt F1}, g}_K({\boldsymbol{x}})&=  \frac{1}{2} \left(\overline{\boldsymbol{u}}_K\cdot \overline{\boldsymbol{F}^g}_K + \overline{\rho\bu}_K \cdot \overline{\left(\frac{\boldsymbol{F}^g}{\rho}\right)}_K\right), ~\rm{and}\\
  {\mathcal F}^{{\tt F2}, g}_K({\boldsymbol{x}})&= \widetilde{\bu}_K \cdot \overline{\boldsymbol{F}^g}_K.
\end{align}
In figure~\ref{fig:injectioncompare} we compare the injection fields obtained using the two definitions for the top right bubble shown in figure~\ref{fig:hga_r8_representative} (c). 
The filtering scale is set at $K \approx 2K_D$, where the largest differences between the definitions are observed.
The analysis reveals that the Favre-filtered contribution {\tt F2} is localized within the bubble, whereas in the velocity-momentum case {\tt F1}, contributions arise from both the bubble interior and its boundary.
We believe that this discrepancy in the two definitions is the reason for the observed differences.
For $K=K_{\max}$, the (unfiltered) buoyancy injection is from the regions within the bubble. We expect the filtered fields to preserve the same structure.
Therefore, the above discussion suggests that the Favre-filtered energy budget is more appropriate to study buoyancy-driven bubbly flows.
Further details are presented in Appendix \ref{app:all_buoyancy}.

\section{Discussion and conclusion}

In this paper, we compared the scale-by-scale energy transfers obtained using two different definitions of the filtered energy, $\mathcal{E}^{\tt F1}_K = \langle\overline{\bu}_K\cdot\overline{\rho\bu}_K\rangle/2$ (velocity-momentum definition) and $\mathcal{E}^{\tt F2}_K = \langle {\overline{\rho}_K}|\widetilde{\bu}_K|^2\rangle/2$ (Favre definition) for bubbly flows.
For the former definition of energy transfer, we also identify an equivalent KHM relation in terms of the velocity-momentum correlation function \citep{GaltierBanerjee2011,Fabien_2025}. 
There are two main contributions of the present manuscript: (a) we find that the KHM relation {\tt C1} shows excellent agreement with the corresponding filtered scheme {\tt F1}, and (b) we find that the filtered energy definitions {\tt F1} and {\tt F2} lead to different physical pictures for the transfer mechanisms when Atwood number is large.
For the parameters explored in our simulations, the nonlinear advection, surface tension and viscous contributions are found to be qualitatively identical for the two definitions.
However, the buoyancy and baropycnal (pressure) contributions are found to be qualitatively different across the two schemes. \newline 
Strikingly, the buoyancy contribution in {\tt F1} also involves an inverse transfer part in addition to the expected injection, which to the best of our knowledge has not been reported in the literature before. In the Favre definition, this inverse transfer is via the baropycnal (pressure) term, while the buoyancy term is a pure injection term \citep{PandeyMitraPerlekar2023}. 
We show that for the $\At$ numbers explored in our study, the two definitions can be matched by extracting the inverse transfer part from $\mathcal{F}^{{\tt F1},\,g}_K$ and adding it to the pressure contribution $\mathcal{P}_K^{\tt F1}$ as discussed in Section \ref{sec:reconcile}.
Thus, irrespective of the definition used there is an inverse energy transfer present, but involve different mechanisms. In the following, we address which definition is the most appropriate.
It is physically reasonable to demand that the injection mechanism (here, the buoyancy contribution) only injects energy and does not couple the resolved and residual scales.
Thus, $\mathcal{F}^g_K$ must be bounded above by $\epsilon^g$ for all $K$, as is the case in definition {\tt F2}. Furthermore, an analysis of the spatial distribution of the buoyancy injection shows that it is localized within the bubble for {\tt F2}, while for {\tt F1} we also observe strong contribution from regions in the vicinity of bubble interface.
These results show that even for moderate Atwood and Galilei numbers used in our simulations the definition {\tt F2} is the more appropriate definition to study energy transfers in buoyancy-driven bubbly flows.\newline 
We now put our manuscript in perspective of the previous studies that have compared different approaches to the scale-by-scale budget.
\citet{Aluie2013} mathematically proved that the Favre {\tt F2} scheme satisfies the inviscid criterion, that is, the viscous effects do not contaminate the inertial-range dynamics.
\citet{ZhaoAluie_2018} showed numerically that, in contrast to {\tt F2}, in the {\tt F3} and {\tt F4} schemes the viscous effects need not be negligible in the intertial range.
While they did not explicitly check the {\tt F1} definition, one expects that this scheme may also violate the inviscid criterion at large Atwood and Galilei (or Reynolds) numbers. 
Another criticism for the {\tt C1} scheme in context of generic compressible flows was raised in \citet{Eyink_2018}, concerned with the regularization of the terms in the $\Rey\to\infty$ limit.
However, as they themselves point out, this criticism is not valid in the incompressible regime $\bnabla \cdot {\boldsymbol u} =0$, which is of interest to the present study.
Finally, the study by \citet{Kritsuk_Wagner_Norman_2013} on forced compressible homogeneous and isotropic turbulence (CHIT) employed the velocity-momentum KHM relation and observed that as $r\to0$ ($K\to\infty$) in filtering framework) the energy injection monotonically saturates to $\langle\rho\bu\cdot\boldsymbol{a}\rangle$ where $\boldsymbol{a}$ is the large-scale stirring acceleration.
While the injection terms like $\langle \bu'\cdot\boldsymbol{F}\rangle$ are, in general, functions of $\boldsymbol{r}$, \citet{Kritsuk_Wagner_Norman_2013} argue that for their choice of forcing and compressible flow dynamics, the injection terms are essentially constant in the inertial range.  
This follows from their observation thatthe fluctuations in the density field have a very short correlation length.
Crucially, in our study, this assumption is not valid since the bubbles represent large-scale coherent fluctuations of density from the mean value, with length-scale $D\approx L/5$ where $D$ is the typical bubble diameter and $L$ is the size of the box. 
\newline To conclude, we compare and contrast two different definitions for the filtered energy {\tt F1} and {\tt F2} for incompressible bubbly flows.
Previous studies, as discussed above, have found the Favre scheme to be more suitable for studying energy transfers in compressible turbulence.
Our work is complementary to these studies in the context of incompressible buoyancy-driven bubbly flows. 
We show that the Favre filtered definition is a natural choice even for moderate density contrast and Galilei numbers.
Our study highlights the importance of appropriately defining the filtered energy and the fact that care must be taken while interpreting the different transfer mechanisms in buoyancy-driven multiphase flows.
\backsection[Acknowledgements]{We thank Supratik Banerjee for discussions.}

\backsection[Funding]{D. M. and V. P. acknowledge the support of the Swedish
Research Council Grants No. 638-2013-9243 and
No. 2016-05225. Nordita is partially supported by
Nordforsk.
H.N. and P. P. acknowledge support from the
Department of Atomic Energy (DAE), India under
Project Identification No. RTI 4007, and DST (India)
Projects No. MTR/2022/000867.
All the simulations are performed using the HPC facility at TIFR
Hyderabad.
}

\backsection[Declaration of interests]{The authors report no conflict of interest.}

\appendix
\section{The K\'arm\'an-Howarth-Monin Relation}\label{app:khm}
\subsection{Connection between the Filtering and Correlation Approaches}\label{sec:mapping_khm}
For completeness, we now make the connection between the filtering and correlation function approaches explicit.
Consider a correlator $C(\br)=\frac{1}{2}\langle A(\bx)\,B(\bx+\br)\rangle$ of $2\upi$-periodic functions $A$ and $B$.
Let the Fourier representation for $A(\bx) = \sum_{\boldsymbol{k}} \widehat{A}(\boldsymbol{k})e^{i\boldsymbol{k}\cdot\bx}$ and likewise $B(\bx)$.
It follows from Parseval's theorem that,
\begin{equation}
C(\br) = \frac{1}{2} \sum_{\boldsymbol{k}}\widehat{A}^*(\boldsymbol{k})\widehat{B} (\boldsymbol{k})e^{i\boldsymbol{k}\cdot\boldsymbol{r}}.\nonumber
\end{equation}
Due to the orthogonality of the Fourier basis, the above expression allows us to immediately write the co-spectrum of $\widehat{A}$ and $\widehat{B}$ in terms of the correlation function as
\begin{equation}
    \langle C(\boldsymbol{r})e^{-i\boldsymbol{k}\cdot\boldsymbol{r}}\rangle_{\boldsymbol{r}} = \frac{1}{2} \widehat{A}^*(\boldsymbol{k})\widehat{B}(\boldsymbol{k}).
\end{equation}
Now, the shell-averaged (isotropic) correlator is defined as,
\begin{equation}
C(r) \equiv \frac{1}{4\upi} \int {\rm d}\Omega\, C(\br) = \frac{1}{2}\sum_{\boldsymbol{k}} \widehat{A}^*(\boldsymbol{k})\widehat{B} (\boldsymbol{k})\frac{\sin(kr)}{kr},\nonumber
\end{equation}
with $\Omega$ being the solid angle.
We can rewrite the right-hand-side above using the spectral energy density $G(\boldsymbol{k}) = \frac{1}{4}(\widehat{A}^*(\boldsymbol{k})\widehat{B} (\boldsymbol{k})+c.c.)$ where $c.c.$ denotes the complex conjugate to get,
\begin{align}
C(r) = \sum_{\boldsymbol{k}} G(\boldsymbol{k})  \frac{\sin (kr)}{kr} = \sum_k G(k)\frac{\sin (kr)}{kr}.\label{eq:WienerKhnichin}
\end{align}
In the second equality above we have done angular integration in $\boldsymbol{k}$-space and $G(k)$ is the energy contained in $k$-shell.
The expression above \eqref{eq:WienerKhnichin} allows one to move between the filtered quantities and correlation functions.
For open boundaries, analogous relation holds with Fourier transform instead of Fourier series, with $\sum\to\int{\rm d} k$.
\subsection{Derivation of the K\'arm\'an-Howarth-Monin Relation}\label{app:deriv_khm}
Our derivation for the KHM relation with the velocity-momentum correlator (definition {\tt C1}) follows closely the presentation given in \citet{GaltierBanerjee2011}.
We first derive the time evolution equation for $R(\boldsymbol{r}) = \frac{1}{2}\langle\boldsymbol{u}'\cdot\rho\boldsymbol{u}\rangle$
Starting from the Navier-Stokes equations,
\begin{equation}
  D_t \rho \bu = -\bnabla p + \mu \nabla^{2} \bu +\boldsymbol{F}^g + \boldsymbol{F}^\sigma,
\end{equation}
and taking the dot product of the above equation with $\bu'$ we get, 
\begin{equation}
 \bu' \cdot \partial_t \rho \bu + \bu' \cdot (\bnabla \cdot \rho \bu \bu) = -\bu'\cdot \bnabla p + \mu \bu'\cdot \nabla^{2} \bu + \bu'\cdot \boldsymbol{F}^g + \bu'\cdot \boldsymbol{F}^\sigma.
\end{equation}
The complementary equation to the above comes from the time evolution equation for the primed velocity field $\bu'$,
\begin{equation}
 \rho\bu \cdot \partial_t \bu' + \rho\bu \cdot (\bnabla' \cdot  \bu' \bu') = -\rho\bu\cdot \frac{\bnabla' p'}{\rho'} + \mu \rho \bu\cdot \frac{1}{\rho'}\nabla'^{2} \bu' + \rho\bu\cdot \frac{1}{\rho'}\boldsymbol{F}^g \,' + \rho \bu\cdot \frac{1}{\rho'}\boldsymbol{F}^\sigma \,'.
\end{equation}
Note that for any function $f(\boldsymbol{x}+\boldsymbol{r})=f'$, we have $\nabla'f'=\nabla f'=\nabla_r f'$ by chain rule. Here, $\nabla_i=\partial/\partial x_i$, $\nabla'_i=\partial/\partial x'_i$ and $\nabla_{ri}=\partial/\partial r_i$.
Adding the above two equations, we have, 
\begin{eqnarray}
   \partial_t (\rho\bu\cdot \bu') &=& - \bu' \cdot (\bnabla \cdot \rho \bu \bu) -  
 \rho\bu \cdot (\bnabla \cdot  \bu' \bu') -\bu'\cdot \bnabla p-\rho\bu\cdot \frac{\bnabla p'}{\rho'}  \\ 
 &+& \mu \bu'\cdot \nabla^{2} \bu +  \mu \rho \bu\cdot \frac{1}{\rho'}\nabla^{2} \bu' + \bu'\cdot \boldsymbol{F} + \rho\bu\cdot \frac{1}{\rho'}\boldsymbol{F}'.
\end{eqnarray}
Here, we are using $\boldsymbol{F}$ as shorthand for $\boldsymbol{F}^g + \boldsymbol{F}^\sigma$. \newline 
We add the above equation the time evolution equation for $R(-\boldsymbol{r})$ and define the scale energy $\mathcal{R}(\boldsymbol{r})$ as,
\begin{equation}
    \mathcal{R} = \frac{1}{4} \langle \rho\bu\cdot\bu' + \rho'\bu'\cdot\bu \rangle.
\end{equation}
In the limit $r\to0$, the above definition of scale energy reduces to the unfiltered kinetic energy.
Any generic term $\boldsymbol{Q}$ that appears on the right-hand-side of the Navier-Stokes equation (for example the dissipation term $\mu\nabla^2 \bu$) has a contribution $\mathcal{Q}$ to the KHM relation that can be written as follows,
\begin{equation}\label{eq:qcontri}
    4 \mathcal{Q} = \left \langle \bu' \cdot \boldsymbol{Q} + \rho \bu \cdot \frac{\boldsymbol{Q}'}{\rho'} + \bu \cdot \boldsymbol{Q}' + \rho' \bu' \cdot \frac{\boldsymbol{Q}}{\rho} \right \rangle.
\end{equation}
There are essentially two contributions to the right-hand-side above, corresponding to terms of type velocity dotted with force, and momentum dotted with acceleration.
We can also rewrite the above expression in terms of increments. Let $\delta$ represent increments in any quantity $\psi$, defined as,
\begin{equation}
    \delta \psi(\boldsymbol{x},\,\boldsymbol{r}) = \psi (\boldsymbol{x}+\boldsymbol{r}) - \psi (\boldsymbol{x}).
\end{equation}
It is then straightforward to check, 
\begin{equation}
    \left\langle\delta \bu \cdot \delta \boldsymbol{Q}\right\rangle = \left\langle\bu'\cdot \boldsymbol{Q}' + \bu\cdot \boldsymbol{Q} - \bu'\cdot \boldsymbol{Q} - \bu\cdot \boldsymbol{Q}'\right \rangle.
\end{equation}
The first two terms are identical after averaging over the entire box, and the last two terms correspond to the first and third term of eq. \eqref{eq:qcontri} (with an extra minus sign here).
Therefore, we can write, 
\begin{equation}
     \mathcal{Q} = \langle \bu\cdot \boldsymbol{Q}\rangle - \frac{1}{4} \left[\left \langle \delta \bu \cdot \delta \boldsymbol{Q} \right\rangle + \left \langle \delta (\rho \bu)\cdot \delta \left(\frac{\boldsymbol{Q}}{\rho}\right) \right\rangle \right].
\end{equation}
By writing the $\mathcal{Q}$ contribution like this, we relate the large-scale contributions (left-side) with the \textit{residual} contributions (right-side in brackets).
Further, the nonlinear term can be written as divergence of a third-order structure function,
\begin{equation}
    4 \mathcal{N} = \left\langle -\bu' \cdot (\bnabla \cdot \rho \bu \bu) - 
 \rho\bu \cdot (\bnabla \cdot  \bu' \bu') - \bu \cdot (\bnabla \cdot \rho' \bu' \bu') - 
 \rho'\bu' \cdot (\bnabla \cdot  \bu \bu)\right\rangle 
\end{equation}
Consider now the following expression, 
\begin{equation}
    \bnabla_r\cdot \langle [\delta (\rho \bu)\cdot \delta \bu \delta \bu ]\rangle.
\end{equation}
The expansion of the above term gives us 8 terms, 
\begin{eqnarray}
\bnabla_r\cdot \langle [\delta (\rho \bu)\cdot \delta \bu \delta \bu ]\rangle &=& \bnabla_r \cdot \langle  [ \rho'\bu'\cdot \bu'\bu'  \nonumber  \\
&-&\rho'\bu'\cdot \bu'\bu - \rho'\bu'\cdot \bu\bu' - \rho\bu\cdot \bu'\bu' \nonumber\\
&+&\rho'\bu'\cdot \bu\bu + \rho\bu\cdot \bu'\bu + \rho\bu\cdot \bu\bu'   \nonumber\\
&-&\rho\bu\cdot \bu\bu] \rangle
\end{eqnarray}
Note that the first and last terms cancel exactly after averaging.
Terms 2 and 7 also vanish because of incompressiblity constraint.
The remaining terms can be seen to be exactly $4\mathcal{N}$, by noting that,
\begin{equation}
     \langle \bnabla_r \cdot(\rho'\bu') \cdot \bu\bu \rangle = \langle \bnabla \cdot(\rho'\bu') \cdot \bu\bu \rangle = -\langle \rho'\bu' \cdot \bnabla \cdot\bu\bu \rangle,
\end{equation}
hence, finally,
\begin{equation}
    \mathcal{N} = \frac{1}{4}\bnabla_r \cdot \langle \delta(\rho\bu)\cdot\delta\bu\,\delta\bu\rangle.
\end{equation}
Here the divergence operator acts on the scale-space coordinates $r$, not the physical space coordinates.
The K\'arm\'an-Howarth-Monin relation for incompressible variable-density flows then is,
\begin{equation}
    \partial_t \mathcal{R}(\boldsymbol{r}) = \mathcal{N}(\boldsymbol{r}) + \mathcal{D}(\boldsymbol{r}) + \mathcal{P}(\boldsymbol{r}) + \mathcal{F}^g(\boldsymbol{r}) + \mathcal{F}^\sigma(\boldsymbol{r}),
\end{equation}
with, $\mathcal{R}(\boldsymbol{r}) = \frac{1}{4} \langle \rho\bu\cdot\bu' + \rho'\bu'\cdot\bu \rangle$. The shell-average of this quantity is $\mathcal{R}(r)$, which represents the cumulative energy in eddies larger than $r$.
In the above derivation, we did not assume either homogeneity or isotropy.
The above derivation can also be derived by interpreting $\langle(\ldots)\rangle$ as ensemble averaging if the flow is assumed to be homogeneous, as is the approach in \citet{Frisch_1995}.

\section{Buoyancy Contribution in {\tt F1-F4} Definitions}\label{app:all_buoyancy}
\begin{figure}
\centering
  \includegraphics[width=0.47\linewidth]{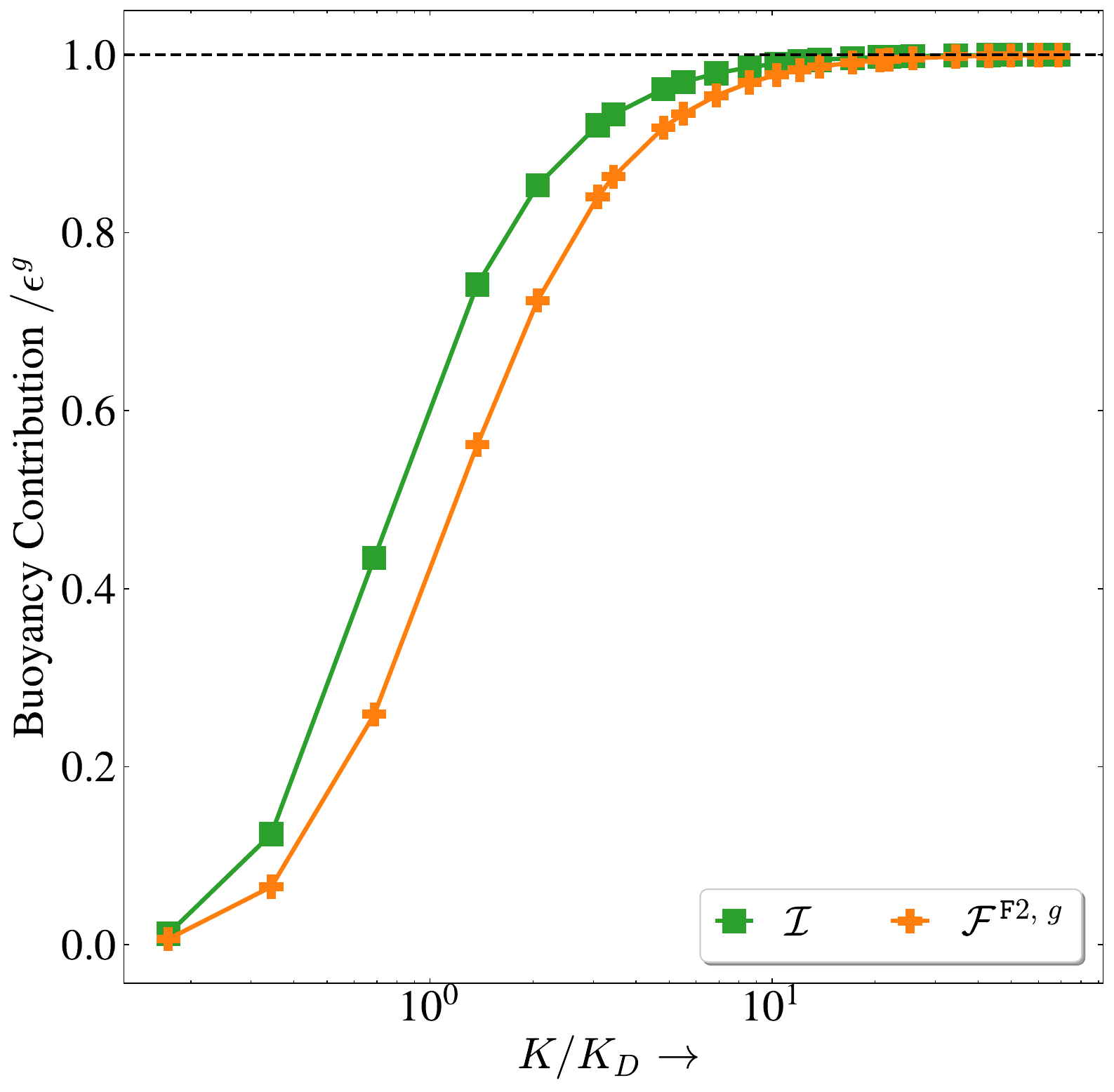}
  \includegraphics[width=0.481\linewidth]{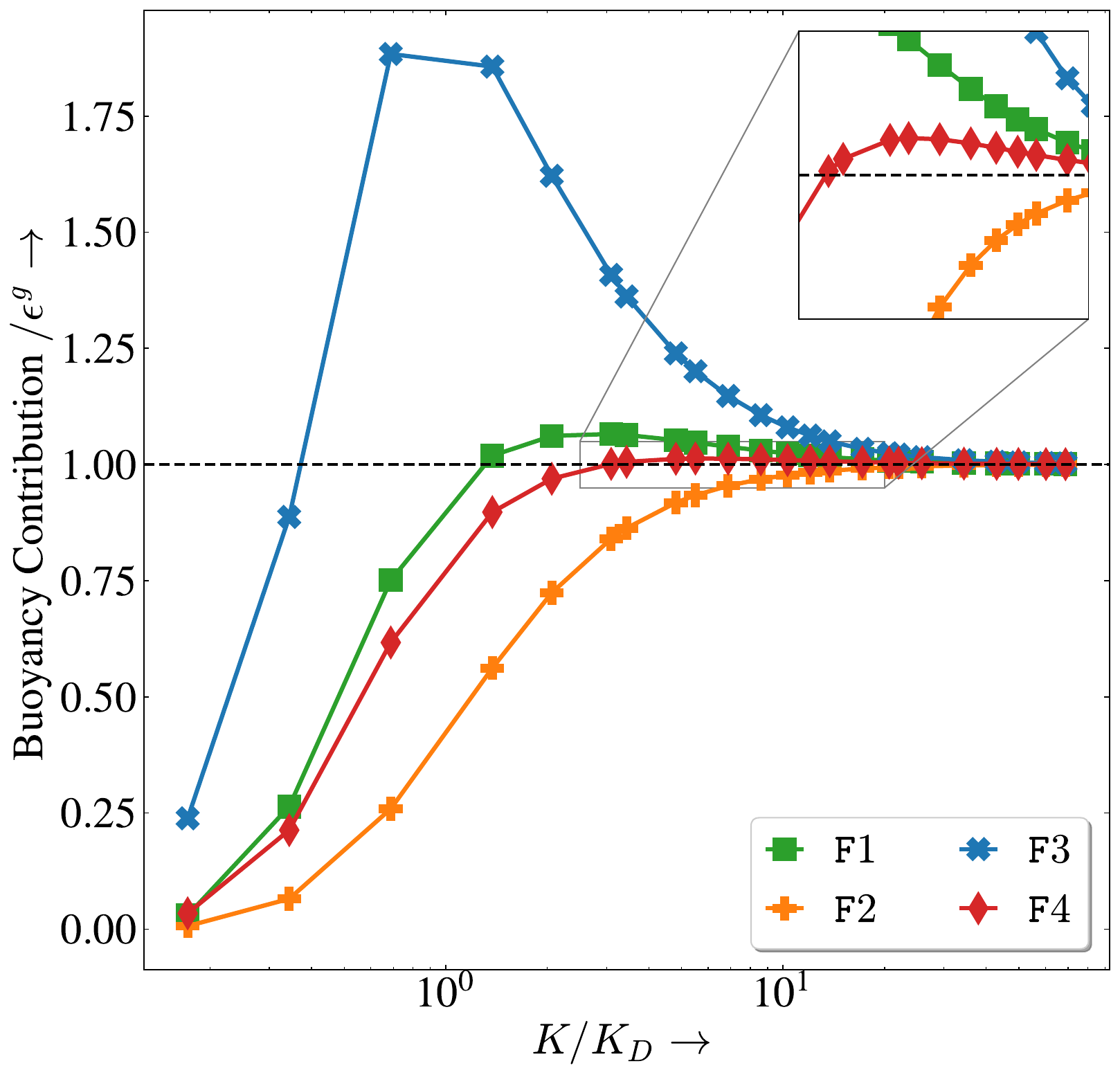}
\caption{(a) Comparison of the monotonic part of buoyancy contribution $\mathcal{I}_K$ in ${\tt F1}$ and the buoyancy contribution $\mathcal{F}^{{\tt F2},\, g}_K$ (left) and (b) the buoyancy contribution for definitions {\tt F1-F4} (right).
Both plots are for run {\tt R2}.
\label{fig:density_buoyancy_compare}}
\end{figure}

We expand here on buoyancy contributions in all the definitions {\tt F1-F4} as presented in Section \ref{sec:ScaleByScale}.
We note that the first (monotonic) part of buoyancy contribution in {\tt F1}, that is, $\mathcal{I}_K$ \eqref{eq:F1grav_break} is close to the Favre contribution at all filtering wavenumbers as shown in figure \ref{fig:density_buoyancy_compare}.
The second part of the buoyancy contribution in {\tt F1} is $\mathcal{J}_K$ \eqref{eq:F1grav_break}.
The form of $\mathcal{J}_K$ is similar to the buoyancy contribution in {\tt F2} except that it has $\overline{(1/\rho)}_K$ instead of $1/\overline{\rho}_K$ in the latter \eqref{eq:f2_contri}.
By Jensen's inequality, $\overline{(1/\rho)}_K\ge1/\overline{\rho}_K$ (point-wise) for all $K$, when filtering is interpreted as a local averaging operation \citep{Aluie2013}.
Indeed, at the interface, $\overline{(1/\rho)}_K\sim1/\rho_B \gg1/\overline{\rho}_K\sim1/\rho_L$ in the limit $\At\to1$. 
Consequently, in comparison to {\tt F2}, the interface is weighted more in the buoyancy contribution for definition {\tt F1}.
Therefore, the difference between the two definitions is expected to become larger as the Atwood number becomes larger. \newline 
Now we consider the four definitions {\tt F1-F4} as defined in Section \ref{sec:ScaleByScale}. 
A similar version of the above argument suggests that the buoyancy contribution $\mathcal{F}^g_K$ in {\tt F1}, {\tt F3} and {\tt F4} definitions is a non-monotonic function of the filtering wave-number $K$, and becomes larger than $\epsilon^g$ for some $K$ in the $\At\to1$ limit.
We verify this in figure \ref{fig:density_buoyancy_compare} for run {\tt R2}.
These results indicate that {\tt F2} definition for the scale-dependent kinetic energy is the most suitable for multi-phase flows in the large Atwood number limit, irrespective of the Galilei number.

\end{document}

%% file: definitions.tex
\definecolor{RoyalBlue}{HTML}{007399}
\definecolor{Seagreen}{HTML}{009999}
\definecolor{crimsonred}{HTML}{990000}
\definecolor{Purple}{HTML}{66032C}

\newcommand{\REM}[1]{{{}}}

\DeclareMathAlphabet\mathbfcal{OMS}{cmsy}{b}{n}

\def \mR {\mathcal{R}}

\def\bu {\boldsymbol{u}}

\def\br {\boldsymbol{r}}

\def \bx {\boldsymbol{x}}

\def \Rey  {\mbox{Re}_{\lambda}}

\def \Ga  {\mbox{Ga}}
\def \At  {\mbox{At}}
\def \Bo  {\mbox{Bo}}

\newcommand{\avg}[1]{\left\langle{#1}\right\rangle}






%% file: revised_new_manuscript.bbl
\begin{thebibliography}{59}
\expandafter\ifx\csname natexlab\endcsname\relax\def\natexlab#1{#1}\fi
\def\au#1{#1} \def\ed#1{#1} \def\yr#1{#1}\def\at#1{#1}\def\jt#1{\textit{#1}} \def\bt#1{#1}\def\bvol#1{\textbf{#1}} \def\vol#1{#1} \def\pg#1{#1} \def\publ#1{#1}\def\arxiv#1{#1}\def\org#1{#1}\def\st#1{\textit{#1}}

\bibitem[Aluie(2011)]{Aluie_2011}
{\sc \au{Aluie, Hussein}} \yr{2011}  \at{Compressible turbulence: The cascade and its locality}.  \jt{Phys. Rev. Lett.}  \bvol{106},  \pg{174502}.

\bibitem[Aluie(2013)]{Aluie2013}
{\sc \au{Aluie, Hussein}} \yr{2013}  \at{Scale decomposition in compressible turbulence}.  \jt{Physica D: Nonlinear Phenomena}  \bvol{247}~(1),  \pg{54--65}.

\bibitem[Aniszewski {\em et~al.\/}(2021)Aniszewski, Arrufat, Crialesi-Esposito, Dabiri, Fuster, Ling, Lu, Malan, Pal, Scardovelli, Tryggvason, Yecko \& Zaleski]{Aniszewski2021}
{\sc \au{Aniszewski, W.}, \au{Arrufat, T.}, \au{Crialesi-Esposito, M.}, \au{Dabiri, S.}, \au{Fuster, D.}, \au{Ling, Y.}, \au{Lu, J.}, \au{Malan, L.}, \au{Pal, S.}, \au{Scardovelli, R.}, \au{Tryggvason, G.}, \au{Yecko, P.} \& \au{Zaleski, S.}} \yr{2021}  \at{Parallel, robust, interface simulator (paris)}.  \jt{Comput. Phys. Commun.}  \bvol{263},  \pg{107849}.

\bibitem[Arun {\em et~al.\/}(2021)Arun, Sameen, Srinivasan \& Girimaji]{Arun_Sameen_Srinivasan_Girimaji_2021}
{\sc \au{Arun, S.}, \au{Sameen, A.}, \au{Srinivasan, Balaji} \& \au{Girimaji, Sharath~S.}} \yr{2021}  \at{Scale-space energy density function transport equation for compressible inhomogeneous turbulent flows}.  \jt{J. Fluid. Mech.}  \bvol{920},  \pg{A31}.

\bibitem[Balachandar \& Eaton(2010)]{Balchandar_2010}
{\sc \au{Balachandar, S.} \& \au{Eaton, John~K.}} \yr{2010}  \at{Turbulent dispersed multiphase flow}.  \jt{Annu. Rev. Fluid Mech.}  \bvol{42}~(Volume 42, 2010),  \pg{111--133}.

\bibitem[Banerjee \& Kritsuk(2017)]{BanerjeeKritsuk_2017}
{\sc \au{Banerjee, Supratik} \& \au{Kritsuk, Alexei~G.}} \yr{2017}  \at{Exact relations for energy transfer in self-gravitating isothermal turbulence}.  \jt{Phys. Rev. E}  \bvol{96},  \pg{053116}.

\bibitem[Bunner \& Tryggvason(2002)]{BUNNER_TRYGGVASON_2002}
{\sc \au{Bunner, Bernard} \& \au{Tryggvason, Grétar}} \yr{2002}  \at{Dynamics of homogeneous bubbly flows part 1. rise velocity and microstructure of the bubbles}.  \jt{J. Fluid. Mech.}  \bvol{466},  \pg{17–52}.

\bibitem[{Chassaing}(1985)]{CHassaing1985}
{\sc \au{{Chassaing}, P.}} \yr{1985}  \at{{An alternative formulation of the equations of turbulent motion for a fluid of variable density}}.  \jt{Journal de Mecanique Theorique et Appliquee}  \bvol{4}~(3),  \pg{375--389}.

\bibitem[Clift {\em et~al.\/}(2005)Clift, Grace \& Weber]{clift2005bubbles}
{\sc \au{Clift, R.}, \au{Grace, J.R.} \& \au{Weber, M.E.}} \yr{2005} {\em Bubbles, Drops, and Particles\/}. {\em Dover Civil and Mechanical Engineering Series\/} .  \publ{Dover Publications}.

\bibitem[Crialesi-Esposito {\em et~al.\/}(2022)Crialesi-Esposito, Rosti, Chibbaro \& Brandt]{Brandt_2022}
{\sc \au{Crialesi-Esposito, Marco}, \au{Rosti, Marco~Edoardo}, \au{Chibbaro, Sergio} \& \au{Brandt, Luca}} \yr{2022}  \at{Modulation of homogeneous and isotropic turbulence in emulsions}.  \jt{J. Fluid. Mech.}  \bvol{940},  \pg{A19}.

\bibitem[Dodd \& Ferrante(2016)]{Dodd_Ferrante_2016}
{\sc \au{Dodd, Michael~S.} \& \au{Ferrante, Antonino}} \yr{2016}  \at{On the interaction of taylor length scale size droplets and isotropic turbulence}.  \jt{J. Fluid. Mech.}  \bvol{806},  \pg{356–412}.

\bibitem[Domaradzki \& Carati(2007)]{DomaradzkiCarati_2007}
{\sc \au{Domaradzki, J.~Andrzej} \& \au{Carati, Daniele}} \yr{2007}  \at{A comparison of spectral sharp and smooth filters in the analysis of nonlinear interactions and energy transfer in turbulence}.  \jt{Phys. Fluids}  \bvol{19}~(8),  \pg{085111},  \arxiv{arXiv: https://pubs.aip.org/aip/pof/article-pdf/doi/10.1063/1.2760281/13636227/085111\_1\_online.pdf}.

\bibitem[Esmaeeli \& Tryggvason(1998)]{ESMAEELI_TRYGGVASON_1998}
{\sc \au{Esmaeeli, Asghar} \& \au{Tryggvason, Grétar}} \yr{1998}  \at{Direct numerical simulations of bubbly flows. part 1. low reynolds number arrays}.  \jt{J. Fluid. Mech.}  \bvol{377},  \pg{313–345}.

\bibitem[Esmaeeli \& Tryggvason(1999)]{ESMAEELI_TRYGGVASON_1999}
{\sc \au{Esmaeeli, Asghar} \& \au{Tryggvason, Grétar}} \yr{1999}  \at{Direct numerical simulations of bubbly flows part 2. moderate reynolds number arrays}.  \jt{J. Fluid. Mech.}  \bvol{385},  \pg{325–358}.

\bibitem[Eyink \& Drivas(2018)]{Eyink_2018}
{\sc \au{Eyink, Gregory~L.} \& \au{Drivas, Theodore~D.}} \yr{2018}  \at{Cascades and dissipative anomalies in compressible fluid turbulence}.  \jt{Phys. Rev. X}  \bvol{8},  \pg{011022}.

\bibitem[Favre(1965)]{Favre_1965}
{\sc \au{Favre, A.~J.}} \yr{1965}  \at{The equations of compressible turbulent gases}.  \jt{Annual Summary Report} .

\bibitem[Frisch(1995)]{Frisch_1995}
{\sc \au{Frisch, Uriel}} \yr{1995} {\em Turbulence: The Legacy of A. N. Kolmogorov\/}.  \publ{Cambridge University Press}.

\bibitem[Galtier \& Banerjee(2011)]{GaltierBanerjee2011}
{\sc \au{Galtier, S\'ebastien} \& \au{Banerjee, Supratik}} \yr{2011}  \at{Exact relation for correlation functions in compressible isothermal turbulence}.  \jt{Phys. Rev. Lett.}  \bvol{107},  \pg{134501}.

\bibitem[Graham {\em et~al.\/}(2010)Graham, Cameron \& Schüssler]{Graham_2010}
{\sc \au{Graham, Jonathan~Pietarila}, \au{Cameron, Robert} \& \au{Schüssler, Manfred}} \yr{2010}  \at{Turbulent small-scale dynamo action in solar surface simulations}.  \jt{Astrophys. J.}  \bvol{714}~(2),  \pg{1606}.

\bibitem[Hamba(2022)]{Hamba_2022}
{\sc \au{Hamba, Fujihiro}} \yr{2022}  \at{Scale-space energy density for inhomogeneous turbulence based on filtered velocities}.  \jt{J. Fluid. Mech.}  \bvol{931},  \pg{A34}.

\bibitem[Hellinger {\em et~al.\/}(2021{\natexlab{{\em a\/}}})Hellinger, Papini, Verdini, Landi, Franci, Matteini \& Montagud-Camps]{Hellinger_2021b}
{\sc \au{Hellinger, Petr}, \au{Papini, Emanuele}, \au{Verdini, Andrea}, \au{Landi, Simone}, \au{Franci, Luca}, \au{Matteini, Lorenzo} \& \au{Montagud-Camps, Victor}} \yr{2021{\natexlab{{\em a\/}}}}  \at{Spectral transfer and kármán–howarth–monin equations for compressible hall magnetohydrodynamics}.  \jt{Astrophys. J.}  \bvol{917}~(2),  \pg{101}.

\bibitem[Hellinger {\em et~al.\/}(2020)Hellinger, Verdini, Landi, Franci, Papini \& Matteini]{Hellinger_2020}
{\sc \au{Hellinger, Petr}, \au{Verdini, Andrea}, \au{Landi, Simone}, \au{Franci, Luca}, \au{Papini, Emanuele} \& \au{Matteini, Lorenzo}} \yr{2020} On cascade of kinetic energy in compressible hydrodynamic turbulence,  \arxiv{arXiv: 2004.02726}.

\bibitem[Hellinger {\em et~al.\/}(2021{\natexlab{{\em b\/}}})Hellinger, Verdini, Landi, Papini, Franci \& Matteini]{Hellinger_2021a}
{\sc \au{Hellinger, Petr}, \au{Verdini, Andrea}, \au{Landi, Simone}, \au{Papini, Emanuele}, \au{Franci, Luca} \& \au{Matteini, Lorenzo}} \yr{2021{\natexlab{{\em b\/}}}}  \at{Scale dependence and cross-scale transfer of kinetic energy in compressible hydrodynamic turbulence at moderate reynolds numbers}.  \jt{Phys. Rev. Fluids}  \bvol{6},  \pg{044607}.

\bibitem[Hill(2001)]{HILL_2001}
{\sc \au{Hill, Reginald~J.}} \yr{2001}  \at{Equations relating structure functions of all orders}.  \jt{J. Fluid. Mech.}  \bvol{434},  \pg{379–388}.

\bibitem[Innocenti {\em et~al.\/}(2021)Innocenti, Jaccod, Popinet \& Chibbaro]{Innocenti_Jaccod_Popinet_Chibbaro_2021}
{\sc \au{Innocenti, Alessio}, \au{Jaccod, Alice}, \au{Popinet, Stéphane} \& \au{Chibbaro, Sergio}} \yr{2021}  \at{Direct numerical simulation of bubble-induced turbulence}.  \jt{J. Fluid. Mech.}  \bvol{918},  \pg{A23}.

\bibitem[Ishihara {\em et~al.\/}(2016)Ishihara, Morishita, Yokokawa, Uno \& Kaneda]{Kaneda_2016}
{\sc \au{Ishihara, Takashi}, \au{Morishita, Koji}, \au{Yokokawa, Mitsuo}, \au{Uno, Atsuya} \& \au{Kaneda, Yukio}} \yr{2016}  \at{Energy spectrum in high-resolution direct numerical simulations of turbulence}.  \jt{Phys. Rev. Fluids}  \bvol{1},  \pg{082403}.

\bibitem[Iyer {\em et~al.\/}(2020)Iyer, Sreenivasan \& Yeung]{Iyer2020}
{\sc \au{Iyer, Kartik~P.}, \au{Sreenivasan, Katepalli~R.} \& \au{Yeung, P.~K.}} \yr{2020}  \at{Scaling exponents saturate in three-dimensional isotropic turbulence}.  \jt{Phys. Rev. Fluids}  \bvol{5},  \pg{054605}.

\bibitem[Kida \& Orszag(1990)]{Kida_1990}
{\sc \au{Kida, Shigeo} \& \au{Orszag, Steven~A.}} \yr{1990}  \at{Energy and spectral dynamics in forced compressible turbulence}.  \jt{J. Sci. Comput.}  \bvol{5}~(2),  \pg{85--125}.

\bibitem[Kritsuk {\em et~al.\/}(2013)Kritsuk, Wagner \& Norman]{Kritsuk_Wagner_Norman_2013}
{\sc \au{Kritsuk, Alexei~G.}, \au{Wagner, Rick} \& \au{Norman, Michael~L.}} \yr{2013}  \at{Energy cascade and scaling in supersonic isothermal turbulence}.  \jt{J. Fluid. Mech.}  \bvol{729},  \pg{R1}.

\bibitem[Kubo {\em et~al.\/}(1991)Kubo, Toda \& Hashitsume]{kubo1991statistical}
{\sc \au{Kubo, Ryogo}, \au{Toda, Morikazu} \& \au{Hashitsume, Natsuki}} \yr{1991} {\em Statistical Physics II: Nonequilibrium Statistical Mechanics\/}, 2nd edn.,  \st{Springer Series in Solid-State Sciences},  \vol{vol.~31}.  \publ{Springer-Verlag Berlin Heidelberg}.

\bibitem[K\"uchler {\em et~al.\/}(2023)K\"uchler, Bewley \& Bodenschatz]{Bodenschatz2023}
{\sc \au{K\"uchler, Christian}, \au{Bewley, Gregory~P.} \& \au{Bodenschatz, Eberhard}} \yr{2023}  \at{Universal velocity statistics in decaying turbulence}.  \jt{Phys. Rev. Lett.}  \bvol{131},  \pg{024001}.

\bibitem[Lai {\em et~al.\/}(2018)Lai, Charonko \& Prestridge]{Lai_Charonko_Prestridge_2018}
{\sc \au{Lai, Chris C.~K.}, \au{Charonko, John~J.} \& \au{Prestridge, Katherine}} \yr{2018}  \at{A kármán–howarth–monin equation for variable-density turbulence}.  \jt{J. Fluid. Mech.}  \bvol{843},  \pg{382–418}.

\bibitem[Lance \& Bataille(1991)]{LanceBataille1991}
{\sc \au{Lance, M.} \& \au{Bataille, J.}} \yr{1991}  \at{Turbulence in the liquid phase of a uniform bubbly air–water flow}.  \jt{J. Fluid. Mech.}  \bvol{222},  \pg{95–118}.

\bibitem[Lees \& Aluie(2019)]{LeesAluie_2019}
{\sc \au{Lees, Aarne} \& \au{Aluie, Hussein}} \yr{2019}  \at{Baropycnal work: A mechanism for energy transfer across scales}.  \jt{Fluids}  \bvol{4}~(2).

\bibitem[Ma {\em et~al.\/}(2022)Ma, Hessenkemper, Lucas \& Bragg]{Ma_Hessenkemper_Lucas_Bragg_2022}
{\sc \au{Ma, Tian}, \au{Hessenkemper, Hendrik}, \au{Lucas, Dirk} \& \au{Bragg, Andrew~D.}} \yr{2022}  \at{An experimental study on the multiscale properties of turbulence in bubble-laden flows}.  \jt{J. Fluid. Mech.}  \bvol{936},  \pg{A42}.

\bibitem[Ma {\em et~al.\/}(2025)Ma, Tan, Ni, Hessenkemper \& Bragg]{ma2025}
{\sc \au{Ma, Tian}, \au{Tan, Shiyong}, \au{Ni, Rui}, \au{Hessenkemper, Hendrik} \& \au{Bragg, Andrew~D.}} \yr{2025}  \at{Kolmogorov scaling in bubble-induced turbulence}.  \jt{Phys. Rev. Lett.}  \bvol{134},  \pg{244001}.

\bibitem[Martinez~Mercado {\em et~al.\/}(2010)Martinez~Mercado, Chehata~Gómez, Van~Gils, Sun \& Lohse]{MARTINEZ_2010}
{\sc \au{Martinez~Mercado, Julián}, \au{Chehata~Gómez, Daniel}, \au{Van~Gils, Dennis}, \au{Sun, Chao} \& \au{Lohse, Detlef}} \yr{2010}  \at{On bubble clustering and energy spectra in pseudo-turbulence}.  \jt{J. Fluid. Mech.}  \bvol{650},  \pg{287–306}.

\bibitem[Mathai {\em et~al.\/}(2020)Mathai, Lohse \& Sun]{ChaoSun_2020}
{\sc \au{Mathai, Varghese}, \au{Lohse, Detlef} \& \au{Sun, Chao}} \yr{2020}  \at{Bubbly and buoyant particle–laden turbulent flows}.  \jt{Annu. Rev. Condens. Matter Phys.}  \bvol{11}~(Volume 11, 2020),  \pg{529--559}.

\bibitem[Miura \& Kida(1995)]{Miura_Kida_1995}
{\sc \au{Miura, Hideaki} \& \au{Kida, Shigeo}} \yr{1995}  \at{Acoustic energy exchange in compressible turbulence}.  \jt{Phys. Fluids}  \bvol{7}~(7),  \pg{1732--1742}.

\bibitem[Ni(2024)]{RuiNi2024}
{\sc \au{Ni, Rui}} \yr{2024}  \at{Deformation and breakup of bubbles and drops in turbulence}.  \jt{Annu. Rev. Fluid Mech.}  \bvol{56}~(Volume 56, 2024),  \pg{319--347}.

\bibitem[Pandey {\em et~al.\/}(2022)Pandey, Mitra \& Perlekar]{Pandey_Mitra_Perlekar_2022}
{\sc \au{Pandey, Vikash}, \au{Mitra, Dhrubaditya} \& \au{Perlekar, Prasad}} \yr{2022}  \at{Turbulence modulation in buoyancy-driven bubbly flows}.  \jt{J. Fluid. Mech.}  \bvol{932},  \pg{A19}.

\bibitem[Pandey {\em et~al.\/}(2023)Pandey, Mitra \& Perlekar]{PandeyMitraPerlekar2023}
{\sc \au{Pandey, Vikash}, \au{Mitra, Dhrubaditya} \& \au{Perlekar, Prasad}} \yr{2023}  \at{Kolmogorov turbulence coexists with pseudo-turbulence in buoyancy-driven bubbly flows}.  \jt{Phys. Rev. Lett.}  \bvol{131},  \pg{114002}.

\bibitem[Pandey {\em et~al.\/}(2019)Pandey, Perlekar \& Mitra]{Pandey_Perlekar_Mitra2019}
{\sc \au{Pandey, Vikash}, \au{Perlekar, Prasad} \& \au{Mitra, Dhrubaditya}} \yr{2019}  \at{Clustering and energy spectra in two-dimensional dusty gas turbulence}.  \jt{Phys. Rev. E}  \bvol{100},  \pg{013114}.

\bibitem[Pandey {\em et~al.\/}(2020)Pandey, Ramadugu \& Perlekar]{Pandey_Ramadugu_Perlekar_2020}
{\sc \au{Pandey, Vikash}, \au{Ramadugu, Rashmi} \& \au{Perlekar, Prasad}} \yr{2020}  \at{Liquid velocity fluctuations and energy spectra in three-dimensional buoyancy-driven bubbly flows}.  \jt{J. Fluid. Mech.}  \bvol{884},  \pg{R6}.

\bibitem[Perlekar(2019)]{Perlekar_2019}
{\sc \au{Perlekar, Prasad}} \yr{2019}  \at{Kinetic energy spectra and flux in turbulent phase-separating symmetric binary-fluid mixtures}.  \jt{J. Fluid. Mech.}  \bvol{873},  \pg{459–474}.

\bibitem[Pope(2000)]{Pope_2000}
{\sc \au{Pope, Stephen~B.}} \yr{2000} {\em Turbulent Flows\/}.  \publ{Cambridge University Press}.

\bibitem[Prakash {\em et~al.\/}(2016)Prakash, Mart{\'i}nez~Mercado, van Wijngaarden, Mancilla, Tagawa, Lohse \& Sun]{Prakash_2016}
{\sc \au{Prakash, Vivek~N.}, \au{Mart{\'i}nez~Mercado, J.}, \au{van Wijngaarden, Leen}, \au{Mancilla, E.}, \au{Tagawa, Y.}, \au{Lohse, Detlef} \& \au{Sun, Chao}} \yr{2016}  \at{Energy spectra in turbulent bubbly flows}.  \jt{J. Fluid. Mech.}  \bvol{791},  \pg{174–190}.

\bibitem[Praturi \& Girimaji(2019)]{Girimaji_2019}
{\sc \au{Praturi, Divya~Sri} \& \au{Girimaji, Sharath~S.}} \yr{2019}  \at{Effect of pressure-dilatation on energy spectrum evolution in compressible turbulence}.  \jt{Phys. Fluids}  \bvol{31}~(5),  \pg{055114}.

\bibitem[Ramadugu {\em et~al.\/}(2020)Ramadugu, Pandey \& Perlekar]{Ramadugu_Pandey_Perlekar_2020}
{\sc \au{Ramadugu, Rashmi}, \au{Pandey, Vikash} \& \au{Perlekar, Prasad}} \yr{2020}  \at{Pseudo-turbulence in two-dimensional buoyancy-driven bubbly flows: A dns study}.  \jt{The Eur. Phys. J. E}  \bvol{43}.

\bibitem[Ramirez {\em et~al.\/}(2024)Ramirez, Burlot, Zamansky, Bois \& Risso]{RAMIREZ2024104860}
{\sc \au{Ramirez, Gabriel}, \au{Burlot, Alan}, \au{Zamansky, Rémi}, \au{Bois, Guillaume} \& \au{Risso, Frédéric}} \yr{2024}  \at{Spectral analysis of dispersed multiphase flows in the presence of fluid interfaces}.  \jt{Int. J. Multiph. Flow}  \bvol{177},  \pg{104860}.

\bibitem[Ravisankar \& Zenit(2024)]{Ravisankar_Zenit_2024}
{\sc \au{Ravisankar, Mithun} \& \au{Zenit, Roberto}} \yr{2024}  \at{Velocity fluctuations for bubbly flows at small re}.  \jt{J. Fluid. Mech.}  \bvol{1001},  \pg{A34}.

\bibitem[Riboux {\em et~al.\/}(2010)Riboux, Risso \& Legendre]{RIBOUX_2010}
{\sc \au{Riboux, Guillaume}, \au{Risso, Frédéric} \& \au{Legendre, Dominique}} \yr{2010}  \at{Experimental characterization of the agitation generated by bubbles rising at high reynolds number}.  \jt{J. Fluid. Mech.}  \bvol{643},  \pg{509–539}.

\bibitem[Risso(2018)]{Risso_2018}
{\sc \au{Risso, Frédéric}} \yr{2018}  \at{Agitation, mixing, and transfers induced by bubbles}.  \jt{Annu. Rev. Fluid Mech.}  \bvol{50}~(Volume 50, 2018),  \pg{25--48}.

\bibitem[Sadek \& Aluie(2018)]{Sadek2018}
{\sc \au{Sadek, Mahmoud} \& \au{Aluie, Hussein}} \yr{2018}  \at{Extracting the spectrum of a flow by spatial filtering}.  \jt{Phys. Rev. Fluids}  \bvol{3},  \pg{124610}.

\bibitem[Salvesen {\em et~al.\/}(2013)Salvesen, Beckwith, Simon, O'Neill \& Begelman]{Salvese_2013}
{\sc \au{Salvesen, Greg}, \au{Beckwith, Kris}, \au{Simon, Jacob~B.}, \au{O'Neill, Sean~M.} \& \au{Begelman, Mitchell~C.}} \yr{2013}  \at{Quantifying energetics and dissipation in magnetohydrodynamic turbulence}.  \jt{Monthly Notices of the Royal Astronomical Society}  \bvol{438}~(2),  \pg{1355--1376}.

\bibitem[Schmidt \& Grete(2019)]{SchmidtGrete_2019}
{\sc \au{Schmidt, Wolfram} \& \au{Grete, Philipp}} \yr{2019}  \at{Kinetic and internal energy transfer in implicit large-eddy simulations of forced compressible turbulence}.  \jt{Phys. Rev. E}  \bvol{100},  \pg{043116}.

\bibitem[Thiesset \& Vahé(2025)]{Fabien_2025}
{\sc \au{Thiesset, Fabien} \& \au{Vahé, Jonathan}} \yr{2025}  \at{Scale-by-scale kinetic energy budgets in multiphase turbulence}.  \jt{J. Fluid Mech.}  \bvol{1025},  \pg{A63}.

\bibitem[Wang {\em et~al.\/}(2013)Wang, Yang, Shi, Xiao, He \& Chen]{Wang_2013}
{\sc \au{Wang, Jianchun}, \au{Yang, Yantao}, \au{Shi, Yipeng}, \au{Xiao, Zuoli}, \au{He, X.~T.} \& \au{Chen, Shiyi}} \yr{2013}  \at{Cascade of kinetic energy in three-dimensional compressible turbulence}.  \jt{Phys. Rev. Lett.}  \bvol{110},  \pg{214505}.

\bibitem[Zhao \& Aluie(2018)]{ZhaoAluie_2018}
{\sc \au{Zhao, Dongxiao} \& \au{Aluie, Hussein}} \yr{2018}  \at{Inviscid criterion for decomposing scales}.  \jt{Phys. Rev. Fluids}  \bvol{3},  \pg{054603}.

\end{thebibliography}
